\documentclass{aa}  

\usepackage{xcolor}
\usepackage{graphicx}
\usepackage{txfonts}
\usepackage{siunitx}
\usepackage{natbib}
\usepackage{amsmath}
\usepackage{subfigure}
\usepackage{hyperref}
\hypersetup{
    colorlinks=true,
    citecolor=blue,
    linkcolor=blue,
    urlcolor=blue,
}

\usepackage{orcidlink}
\newcommand{\orcid}[1]{\href{https://orcid.org/#1}
{{\includegraphics[height=8pt]{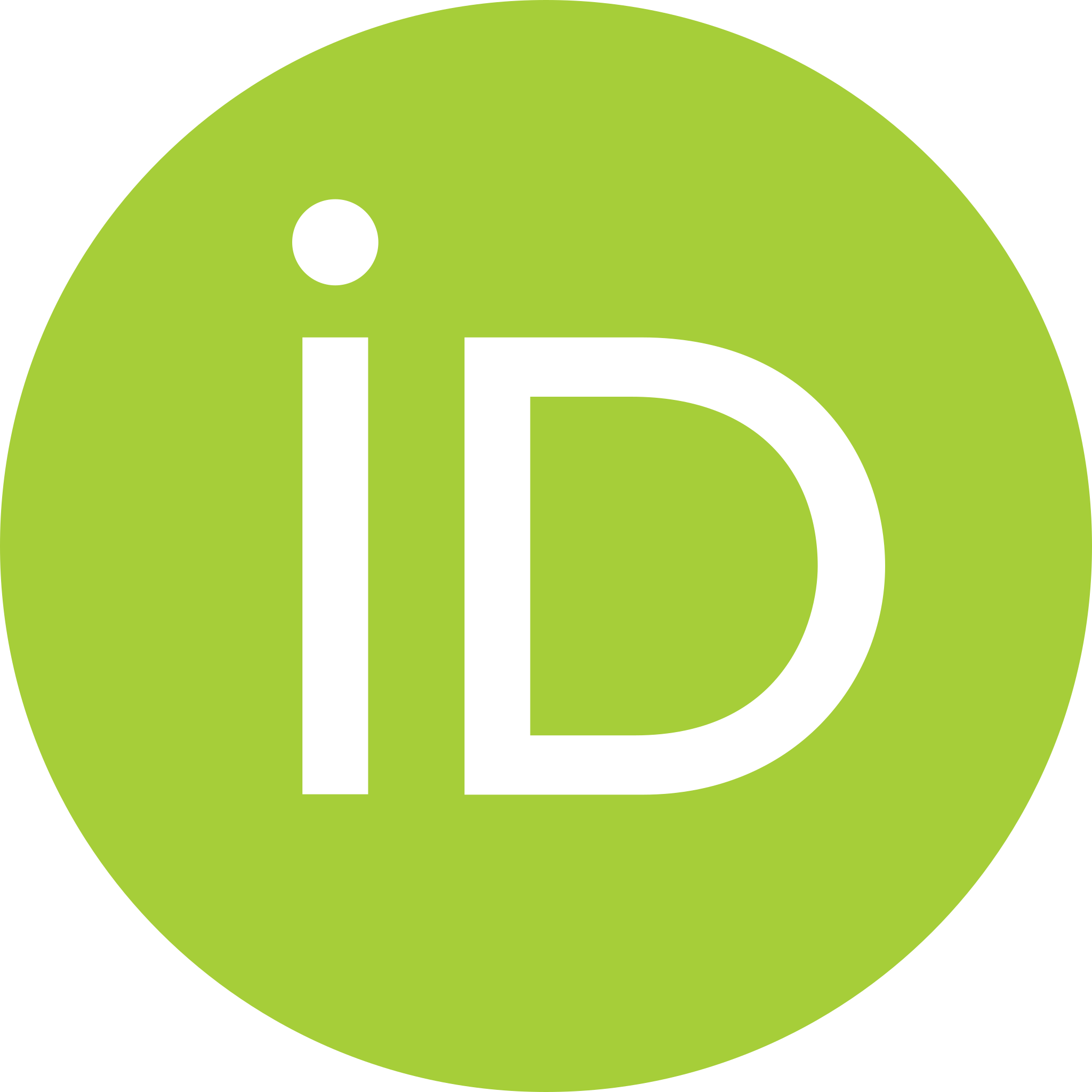}}}}

\def\bhm{M$_{\rm BH}$}
\def\lbhm{log(M$_{\rm BH})$}

\def\lum{L$_{\rm bol}$}
\def\llum{log(L$_{\rm bol}$)}
\def\ledd{$\lambda_{Edd}$}
\def\lledd{log($\lambda_{Edd})$}

\usepackage{ulem}   
\definecolor{mpc}{rgb}{0.2, 0.2, 0.8}
\definecolor{vpc}{rgb}{0.2, 0.8, 0.2}

\begin{document}

   \title{The ensemble broad-frequency power spectrum of Stripe-82 quasars from multiple surveys}

    \author{V. Petrecca\thanks{vincenzo.petrecca@inaf.it}\inst{1,2}\orcid{0000-0002-3078-856X}
          \and
          I. E. Papadakis\inst{3,4}\orcid{0000-0001-6264-140X}
          \and
          M. Paolillo\inst{2,1,5}\orcid{0000-0003-4210-7693}
          \and 
          D. De Cicco\inst{2,1,6}\orcid{0000-0001-7208-5101}
          \and
          F. E. Bauer\inst{7}\orcid{0000-0002-8686-8737}
          \and
          M. I. Carnerero\inst{8}\orcid{0000-0001-5843-5515}
          \and
          C. M. Raiteri\inst{8}\orcid{0000-0003-1784-2784}
          \and
          M. Fatovi\'c \inst{2,1,9}\orcid{0000-0003-1911-4326}
          }

   \institute{INAF - Osservatorio Astronomico di Capodimonte, Salita Moiariello 16,       80131 Napoli, Italy
        \and
        Dipartimento di Fisica    "Ettore Pancini", Università di Napoli Federico II, via Cinthia 9, 80126 Napoli, Italy
        \and
        Department of Physics and Institute of Theoretical and Computational Physics, University of Crete, 71003 Heraklion, Greece
        \and
        Institute of Astrophysics, FORTH, GR-71110 Heraklion, Greece
        \and
        INFN - Sezione di Napoli, via Cinthia 9, 80126, Napoli, Italy
        \and
        Millennium Institute of Astrophysics, Nuncio Monse{\~{n}}or S{\'{o}}tero Sanz 100, Of 104, Providencia, Santiago, Chile
        \and
        Instituto de Alta Investigaci{\'{o}}n, Universidad de Tarapac{\'{a}}, Casilla 7D, Arica, Chile
        \and
        INAF - Osservatorio Astrofisico di Torino, via Osservatorio 20, I-10025 Pino Torinese, Italy
        \and
        Ruđer Bošković Institute, Bijenička Cesta 54, 10000 Zagreb, Croatia
        }

   \date{Received 29 August 2025 / Accepted 17 November 2025}

 
  \abstract
   {Variability is one of the most striking features of quasars, observed at all timescales and throughout the electromagnetic spectrum. The study of variability properties and their correlations with the physical parameters (e.g. black hole mass and accretion rate) provides significant insights into accretion physics. However, the detailed picture and the exact interplay between different emitting regions are not yet clear.}
   {We combine data from Sloan Digital Sky Survey (SDSS), the Panoramic Survey Telescope and Rapid Response System 1 (Pan-STARRS1, PS1), the Zwicky Transient Facility (ZTF), and the Gaia space telescope to constrain the power spectrum of quasars in the Stripe-82 region over a broad frequency range, from $10^{-1}$ to $10^{-3}\ day^{-1}$ in the rest frame.}
   {The light curves of multiple surveys were matched and cross-calibrated to reach $\sim 20$ years in the r band for 4037 quasars. We split the sample into bins of the same black hole mass, accretion rate, and redshift (as a proxy of rest-frame wavelength) and measured the ensemble power spectral density (PSD) in each bin. The power spectra of SDSS, ZTF, and Gaia were measured independently. We did not measure the PSD on PS1 data due to a more erratic cadence, as well as the similarity in terms of baseline compared to the other surveys. However,  we discuss the use of interpolation techniques that eventually enable us to use the data together and probe frequencies lower than $10^{-3}\ day^{-1}$ in the rest frame.}
   {We find significant evidence that the long-term ultraviolet/optical variability of quasars is stationary, as the ensemble PSD estimates from SDSS, Gaia, and ZTF are consistent within the errors despite their originating from different surveys and different years. The PSD shape is consistent with a bending power law with spectral indices of --2.7 and --1 at high and low frequencies. A fit with the model PSD associated with a damped random walk model (spectral indices --2 and 0) is significantly worse. The power spectrum amplitude below the break does not depend on black hole mass, but there is some evidence to support an anti-correlation with the accretion rate. The bending frequency, instead, scales with the black hole mass as $\nu_b \propto M_{BH}^{-0.6\pm0.1}$ and it does not depend on the accretion rate.}
   {}
   

   \maketitle
%
\section{Introduction}
\label{sec:intro}
Active galactic nuclei (AGNs) are among the most energetic sources in the Universe, powered by supermassive black holes (BHs) actively accreting matter at the centre of galaxies (\citealt{Netzer2015}). One of their striking features is variability, observed across the entire electromagnetic spectrum, both in the continuum and the emission lines, with timescales ranging from minutes to years (e.g. \citealt{Ulrich1997}; \citealt{Giveon1999}; \citealt{2021Cackett}). The study of AGN variability is becoming increasingly relevant, especially in the era of large time-domain surveys. Indeed, variability allows us to understand the geometry and physics of the accretion process and has proven to be a valuable selection tool (e.g. \citealt{Butler2011}; \citealt{DeCicco2022,Decicco2025}; \citealt{Savic2023}). Most of the studies focus on continuum X-ray and ultraviolet/optical (UV/optical) variability measured from light curves, which show a stochastic behaviour and directly probe the accretion disc and the X-ray corona around the central source. Although there is debate about the interplay between these two AGN regions, the correlations between variability and AGN properties (e.g. BH mass, accretion rate, wavelength) and the origin of variability itself, there is general consensus that power spectral density functions (PSDs or power spectra) or equivalent measurements (e.g. structure function, SF) display a red-noise trend and are usually modelled with one or more power laws (see \citealp{Paolillo2025} and references therein for a recent review of continuum variability). Optical light curves are usually modelled with a damped random walk (DRW; \citealt{Kelly+09}). This is a first-order continuous-time autoregressive moving average (CARMA) process, which predicts a PSD with a spectral index of --2, flattening to 0 after a damping timescale, $\tau$. Although the model seems to be consistent with light curves of the order of a few years, deviations have been observed on both shorter and longer timescales (e.g. \citealt{Mushotzky+11}; \citealt{Guo+17}), as well as possible biases in the proper recovery of DRW parameters due to sampling effects (e.g. \citealt{Emmanoulopoulos2010}; \citealt{kozlowski+17}). To improve our understanding of AGN continuum variability, we need large samples of light curves that probe multiple timescales and to explore new analysis techniques, such as fitting the data with higher-order CARMA models (e.g. \citealt{yu2022}), unsupervised machine learning analysis (e.g. \citealt{Tachibana2020}), or directly measuring the PSD with no a priori model assumptions. 

In \cite{Petrecca2024}, hereafter P24, we studied a sample of 8042 spectroscopically confirmed quasars (i.e. the most luminous among unobscured AGNs) from the SDSS Stripe-82 region, first selected and analysed by \citet{MacLeod10,MacLeod2012}. Unlike many of the previous works on optical light curves, P24 directly computed the power spectra via traditional Fourier techniques and studied relations between the ensemble PSD and quasar physical properties. They found that variability does not depend on redshift, while both the PSD amplitude and slope depend on the accretion rate, BH mass, and rest-frame wavelength (e.g. Eqs. 16 and 17 in P24). They also discussed the possibility of a universal PSD shape for all quasars, where frequencies scale with the BH mass, while normalisation and slopes are ﬁxed (at any given wavelength and accretion rate). While similar ideas have also been proposed by other studies (e.g. \citealt{Tang2023}; \citealt{Arevalo2024}), limitations in cadence, temporal baseline, and survey coverage have hindered the ability to unambiguously determine the intrinsic PSD shape of all AGN and its dependencies. 

The forthcoming Legacy Survey of Space and Time (LSST; \citealt{Ivezic2019}) expected to start by the end of 2025 at the Vera C. Rubin Observatory will mitigate many of these problems. It will observe the entire southern sky every three-four nights for ten years across \textit{ugrizy} filters and a $5\sigma$ point source depth of $\sim 24.7\ mag_{AB}$ in the r band for a single visit (\citealt{Bianco2022}). In addition, five deep drilling fields (DDF\footnote{\hyperlink{https://survey-strategy.lsst.io/baseline/ddf.html}{survey-strategy.lsst.io/baseline/ddf}}) covering $\sim 60\ deg^2$ will be observed even more intensively in terms of depth and cadence. The LSST is expected to detect tens of millions of AGNs, allowing for a detailed analysis of light curves on multiple timescales, including fast and extreme variability (e.g. \citealt{Raiteri2022}; \citealt{Komossa2024}), and precise time-delay measurements (\citealt{Kovacevic2022}; \citealt{Czerny2023}). However, challenges such as seasonal gaps, uneven sky coverage, and missing data due to other survey-related factors will inevitably persist. There is a general community effort based on simulations and archival data aimed at supplementing our understanding of what to expect from the LSST data, while preparing a machinery of software, techniques, and ancillary datasets ready to work with the first Rubin data releases. Our PSD framework, introduced in P24, is among these methods, but other aspects such as optimising light-curve pre-processing to extract more variability features also need to be inspected. 

This work is a follow-up to the P24 paper. We present an ensemble PSD analysis of quasars combining data from multiple surveys. The starting sample consists of the SDSS Stripe-82 quasars from P24, which  all feature information related to redshift, BH mass (\bhm), luminosity (\lum), and accretion rate (\ledd=$L_{Bol}/L_{Edd}$). The combination of light curves from different facilities allows us to sample a larger frequency range with both shorter and longer timescales, thus having a better constraint on the PSD shape. Moreover, this methodology will be most likely applied in the future to LSST data to extend the observing baseline, but taking advantage of the better cadence and wavelength coverage similar to the SDSS data studied in P24. Finally, we also present a library of calibrated light curves covering $\sim 20$ years, expanding on previous efforts by \cite{Rumbaugh2018}, \cite{Suberlak+21}, and \cite{Stone2022}. 

This paper is organised as follows. In Sect. \ref{sec:data}, we present the data used for our work. Section \ref{sec:PSD_estimate} describes the PSD estimates from multiple surveys, while in Sect. \ref{sec:psd-shape} we discuss the broad-band PSD shape. We investigate how the power spectra depend on BH mass and accretion rate in Sect. \ref{sec:trends} and we report our final considerations in Sect. \ref{sec:discussion}.

\section{Data}
\label{sec:data}

\begin{figure*}
    \centering
    \includegraphics[width=0.49\linewidth]{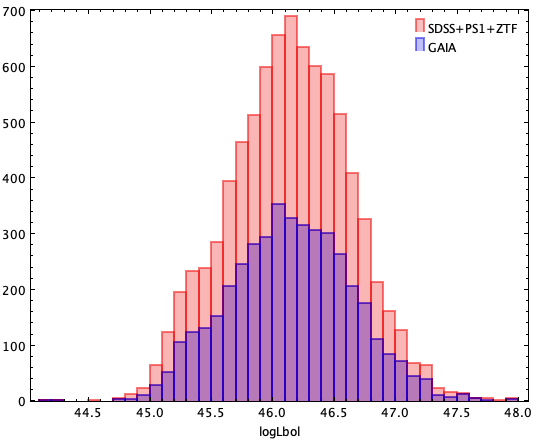}
    \includegraphics[width=0.49\linewidth]{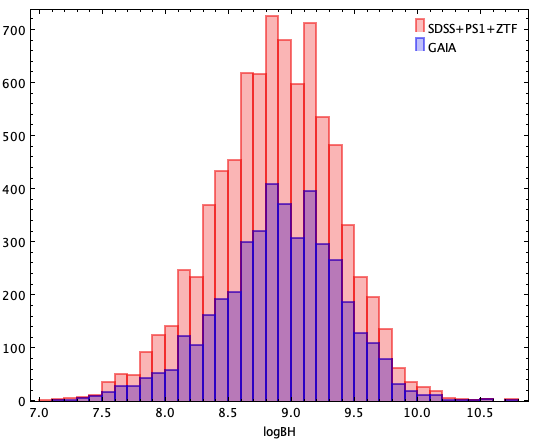}
    \caption{Distributions of bolometric luminosity (left panel) and BH mass (right panel) for the 8042 SDSS+PS1+ZTF quasars (red) and the subset of 4037 sources in common with Gaia DR3 (blue).}
    \label{fig:Gaia_sdss_hist}
\end{figure*}

In an effort to build a legacy dataset of archival light curves in preparation for LSST, we started collecting all publicly available data complementing the SDSS Stripe-82 region already studied for the LSST AGN Science Collaboration Data Challenge (\citealt{yu2022lsstc}; \citealt{Savic2023}) by P24 and others (e.g. \citealt{MacLeod10}; \citealt{kozlowski2016}). As our sample of quasars from Stripe-82 consists of bright point-like sources with mag < 21.5 (see Fig. 1 in P24), it is also possible to recover most of them with shallower and lower resolution surveys. The point-like nature of quasars enables an easier comparison of PSF photometry from different facilities, with minor issues related to the host-galaxy contribution. 

The most recent, ongoing time-domain survey is the ZTF (\citealt{Bellm2019}; \citealt{Graham2019}), which shares many similarities with the future LSST, including alert production and difference image analysis (DIA), although at a lower depth. It started in March 2018, using the Samuel Oschin 48-inch Telescope at the Palomar Observatory to observe the northern equatorial sky (including Stripe-82) in \textit{g,r} bands with a few nights cadence. Additional \textit{i} band observations are also available (although with a more erratic cadence), as well as repeated \textit{g,r} exposures in the same night. Currently, public data releases are updated every two months and include raw CCD-based images and calibrations in FITS format, reference and difference images from DIA, updated objects database, and single-epoch photometry (\citealt{Masci2019}). We used ZTF DR19 to cross-match the object table with the catalogue of SDSS Stripe-82 quasars. The match returned 8250 sources within 0.5 arcsec (mean separation 0.12 arcsec, standard deviation 0.06 arcsec), including all the 8042 quasars used by P24. We downloaded PSF photometry in \textit{g,r} bands up to July 2023 (i.e. the last visits in DR19) for each object with an ad hoc python code querying the IRSA database\footnote{\href{https://irsa.ipac.caltech.edu/data/ZTF/lc/lc_dr19/}{https://irsa.ipac.caltech.edu/data/ZTF/lc/lc\_dr19/}.}. 

Although the ZTF high-cadence data are ideal for studying variability on short timescales, the temporal baseline of the DR19 used in this work is $\sim\!5.5$ years and it does not add any information on timescales longer than those probed with the SDSS light curves (the case is the same when using the most recent data release at the time of writing, which is about seven years, but this could change in the future if the survey progresses and overlaps with the LSST). Furthermore, the close to nine year gap between the two surveys prevents us from simply combining the light curves to constrain even longer times. Filling the gap requires gathering archival data from other facilities, as proposed by \cite{Suberlak+21}; see also Figure 2 of \cite{Paolillo2025} for an updated list. However, the differences between the surveys in terms of sky coverage, photometric systems, sensitivity, and resolution make combining and homogenising the data a non-trivial task; thus, the selection of a good sample is critical to minimise any bias in the analysis. Among recent time-domain surveys that overlap with the Stripe-82 region, we have  Catalina Real Time Survey (CRTS; \citealt{Drake2009}),  Palomar Transient Factory (PTF/iPTF survey; \citealt{Law2009}),  Panoramic Survey Telescope And Rapid Response System Survey (Pan-STARRS-1, or PS1 for short; \citealt{Kaiser2010}; \citealt{Chambers2016}),  All-Sky Automated Survey for Supernovae (ASAS-SN; \citealt{Shappee2014}),  NASA Asteroid Terrestrial-impact Last Alert System (ATLAS; \citealt{Tonry2018}), and the Gaia space telescope (\citealt{Gaia2016}).

In an attempt to use all the available information, we explored the light curves of each survey and reached the same conclusion as \cite{Suberlak+21}, namely, to exclude CRTS and PTF from the analysis. In fact, their shallower photometry and large photometric uncertainties affect the quality of the observed PSDs, whereas the temporal coverage is similar to the deeper PS1 observations (actually CRTS has a better sampling, but this is spoiled by the lower photometric quality and the need to reduce the sample to fewer good sources). Similar considerations also apply to ATLAS and ASAS-SN, whose limiting magnitude is shallower than the typical $\sim 20\ mag_{AB,r}$ of Stripe-82 quasars (see Fig. 1 in P24). On the other hand, we included ZTF data in our analysis, as they provide a fundamental contribution to the high-frequency PSD, as shown in Sect. \ref{sec:ztf_psd}, and cover a different time period. As a further ground-based survey, we included PS1 as well, a wide-field imaging facility that contains measurements in five filters, \textit{grizy}, for $30\ 000$ square degrees of the sky north of declination --30° (including Stripe-82). The data set is composed of roughly $\sim12$ epochs for each filter, from May 2010 to March 2014, publicly available through the DR2\footnote{\href{https://outerspace.stsci.edu/display/PANSTARRS/}{https://outerspace.stsci.edu/display/PANSTARRS/}.} (\citealt{Flewelling2020}). A spatial cross-match with the Stripe-82 quasar catalogue returns all the sources in common with ZTF, but there is still a significant gap of $\sim$ 4-5 years between the observations. 

We also used light curves from the ESA's Gaia satellite, designed for an accurate astrometric, photometric, and spectroscopic map of the Milky Way, therefore exhibiting extremely good resolution and photometric quality, albeit less deep and employing broad-band filters. Having started its operations in late 2014 and  just concluding its nominal mission in early 2025, Gaia is the perfect survey to complement the previous observations. We cross-matched the Stripe-82 quasars with the Gaia DR3 variable AGN catalogue (\citealt{Carnerero2023}), covering 34 months from July 2014 to May 2017. We found a match for 4037 objects. This low number is due to many factors, including the lower magnitude limit ($mag_G <21$) and the fact that Gaia only published light curves for sources detected as variable within the survey. The last limitation is the most severe, considering that AGN variability tends to be larger on longer timescales and a few years might not be enough for a complete selection (e.g. see discussion in \citealt{DeCicco2019}). However, by looking at the quasar physical properties from SDSS+PS1+ZTF and Gaia (Fig. \ref{fig:Gaia_sdss_hist}), we see that they cover the same range of \lum\ and \bhm\ with a similar distribution; thus, the analysis on the limited sample is not expected to introduce any severe bias. Future data releases could allow for the recovery of the light curves for the missing sources and the addition of more epochs, possibly overlapping with ZTF and improving the analysis. 

To summarise, we cross-matched data from SDSS (1998--2008), PS1 (2010--2014), Gaia (2014--2017), and ZTF (2018--2023) to study AGN variability and build a legacy dataset in preparation for the LSST. The ZTF's high cadence allows for a detailed analysis of high-frequency PSDs, while the full combined light curves cover more than 20 years, probing lower frequencies. 
Our main objective is to compute the power spectrum of the quasars in the widest possible frequency range. To do so, the AGN flux from the various catalogues must be cross-calibrated. This is necessary for various reasons. First, the filters used by the SDSS, PS1 and ZTF surveys are not identical (in width and shape). Second, Gaia measurements are made using a single broad filter, which partially overlaps but is not similar to any of the filters used by the other surveys. 
Therefore, it is necessary to cross-calibrate all the data we use to a single waveband (or filter) before using them to calculate the PSD. 

\begin{figure*}
    \centering
    \includegraphics[width=0.8\linewidth]{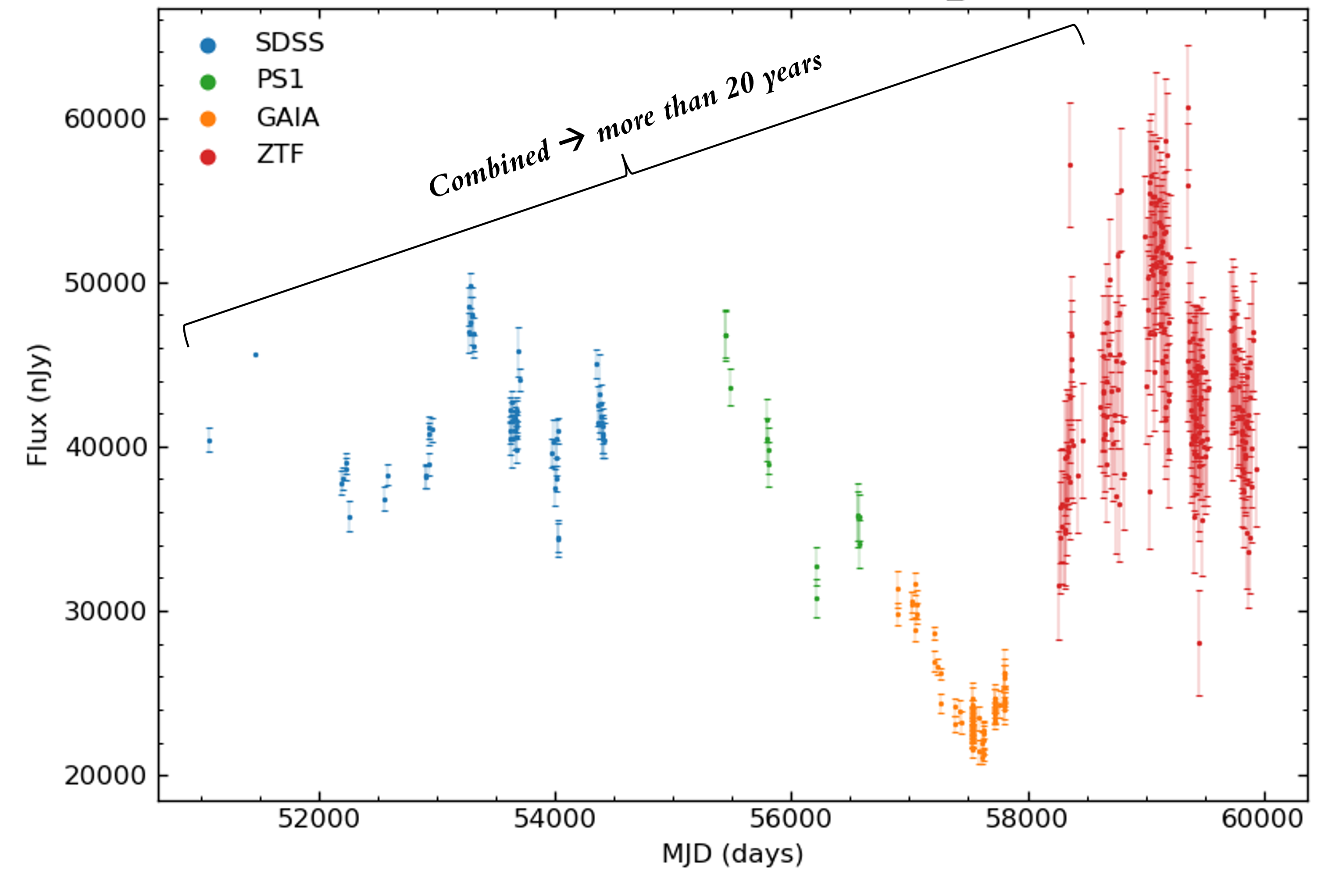}
    \caption{Example of r-band light curve for a Stripe-82 quasar combining SDSS (blue), PS1 (green), Gaia (orange), and ZTF (red) data. Colour corrections and calibrations are given in Appendix \ref{subsec:agn_lc_calib}.}
    \label{fig:agn_lc_comb}
\end{figure*}

In Appendix \ref{subsec:agn_lc_calib}, we explain in detail the process we followed to calibrate all light curves from the various surveys to the SDSS $r-$filter flux, which is the band minimising the corrections. We show an example of such a light curve in Fig. \ref{fig:agn_lc_comb}, highlighting the contributions from the different surveys. The cross-calibrated light curves prepared for this work will be made available to the community.

\section{Estimating the power spectrum.}
\label{sec:PSD_estimate}
As the variability of AGNs is a stochastic process, we would ideally like to calculate the power spectrum on the longest and best sampled light curves as possible for a significant number of sources. In fact, sampling a given time range multiple times for sources with similar properties is the only way to determine the typical variability behaviour and it is the main motivation behind the effort of building a dataset as the one described in Sect. \ref{sec:data}. To study the variability, we used traditional Fourier techniques to estimate the PSD; namely, we used the periodogram as an estimator of the power spectrum, as done in P24. Fourier analysis is fast and reliable, and the periodogram is a well-studied robust estimator of the power spectrum of stationary random processes (e.g. \citealt{Press1978}; \citealt{Priestley1981}; \citealt{vanderklis1988}; and references therein). We can also bin a large number of periodograms \citep[at least 50; see e.g.][]{Papadakis1993} to calculate appropriate estimators to model fit the power spectra using ordinary $\chi^2$ minimisation. However, traditional Fourier analysis requires evenly sampled light curves. 

We tried combining the SDSS+PS1+Gaia+ZTF data using bins of various sizes, but were unable to construct homogeneous, long, and evenly sampled light curves (with a reasonable number of missing points), spanning the whole interval from the start of the SDSS observations to the end of the ZTF data points. This was mainly due to the large gap between SDSS and the Gaia data points and the erratic sampling of quasars by PS1. For that reason, we adopted a different strategy. We used the ZTF data for the determination of the high-frequency power spectrum of the sources, and we kept the SDSS PSD estimates of P24 at low frequencies. To improve the determination of the low-frequency PSD, we calculated additional low-frequency PSD estimates using the individual Gaia and the ZTF light curves appropriately binned, using large bin sizes. We explain in detail the calculation of the high- and low-frequency PSD in the following sub-sections.

\subsection{High-frequency PSD from ZTF}
\label{sec:ztf_psd}
\begin{figure}
    \includegraphics[width=0.99\columnwidth]{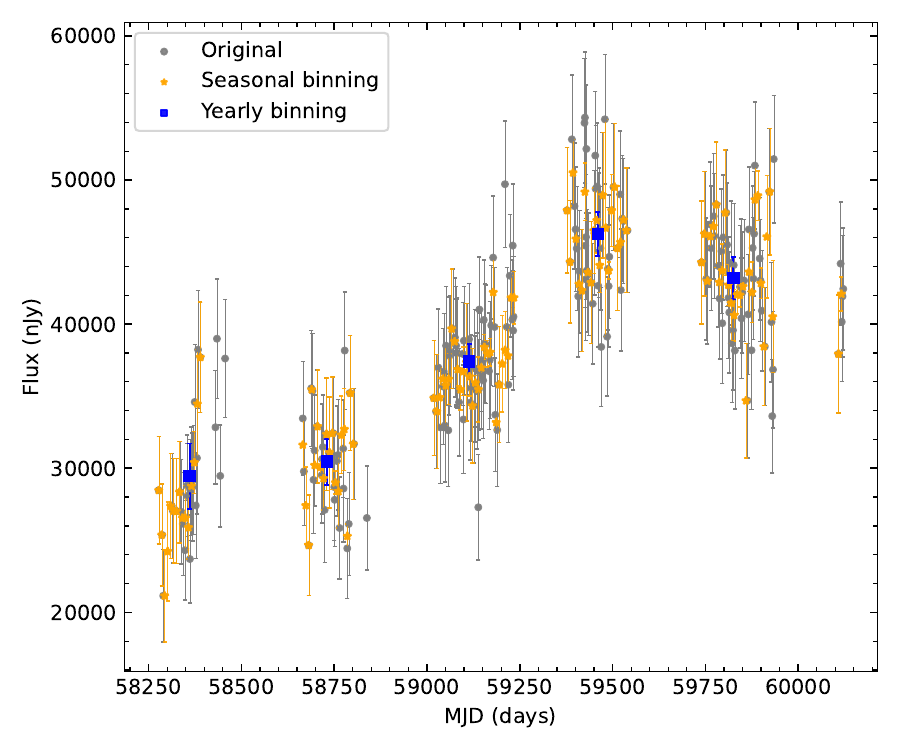}
    \caption{Example light curve for a ZTF quasar in the \textit{r} band. Grey points are the original flux measurements, while yellow stars indicate the binned-interpolated values turning each season into an evenly sampled time series to compute the high-frequency PSD in Sect. \ref{sec:ztf_psd}. The second half of the first season shows data points rejected from the analysis due to the quality cuts applied. Blue squares are yearly-binned points used for the low-frequency PSD in Sect. \ref{sec:combined1}.}
    \label{fig:agn_lc_ztf}
\end{figure}

ZTF light curves from DR19 are composed of five yearly seasons, each with a cadence of a few nights  and about five-six months in length. To measure the PSD on very short timescales, we considered each season separately assuming that they are multiple realisations of the same stationary random process. First, we converted magnitudes into fluxes in units of nJy and obtained an evenly sampled time series for each seasonal segment of the light curve, in the same way as described in P24 (i.e. starting from magnitudes in the AB photometric system with a zero point of 3631 Jy and following a standard procedure for error propagation). We adopted a bin size of 12 days, which allows us to have at least three independent visits in most of the bins. For each bin, we calculated the mean flux and the standard error on the mean (i.e. $\sigma/\sqrt{n}$, where $\sigma$ is the standard deviation and $n$ is the number of points). As is often the case with high-cadence surveys, there are missing observations within each seasonal segment (e.g. due to bad weather, full moon, etc.). We used linear interpolation to fill the gaps. We rejected segments having more than three consecutive gaps (i.e. > 30 days of missing data) and when the interpolated values accounted for more than 15\% of the total number of points in the segment. This choice maximises the amount of data used for the analysis (more than $\sim 80$\% of the seasonal segments for the quasars we considered), while minimising biases introduced from the interpolation procedure, as we show on simulated light curves in Appendix \ref{sec:app_b}. We also tested different rejection thresholds and higher-order interpolation methods, but linear interpolation appears to be the best option for the sort term light curves. An example of the ZTF light curve with the binned data plotted on top of the observations is shown in Fig. \ref{fig:agn_lc_ztf}. 

\subsection{Low-frequency PSD from SDSS, ZTF, and Gaia}
\label{sec:combined1}

In P24, we calculated the power spectrum at three low frequencies from the SDSS data, using six-year-long light curves with a yearly bin size (i.e. we used the last six SDSS observing seasons from the typical light curve in Fig. \ref{fig:agn_lc_comb} as the epochs before were sparser; see P24 for more details). The Gaia light curves are also sufficiently well sampled to be binned with a 180-day window. The resulting light curves do not have (almost any) missing points and can be used to calculate the periodogram at three new independent frequencies, located between the SDSS and ZTF frequencies. Two additional PSD estimates at frequencies within the range sampled by SDSS could also be obtained by using the full ZTF light curves when they are binned with a bin size of one year. Contrary to Gaia and ZTF, the PS1 light curves' sampling is more erratic and makes it difficult to to construct evenly sampled light curves without adding a large number of interpolated points. For that reason, we did not use the PS1 light curves to calculate the power spectrum.

We calculated separate periodograms for each seasonal segment in ZTF light curves, for the full GAIA light curves, and for the full ZTF light curves, as in P24, namely,\begin{equation}
I_{N}(f_j)  =  \frac{2\Delta t}{N\bar{x}^2} \ \Bigg[\sum^{N}_{i=1} [x(t_i)-\bar{x}]e^{-i2\pi f_jt_i}\Bigg]^2,
\label{periodogram}
\end{equation}

\noindent where $N$ is the number of points in the light curve, $x(t_i)$ are the flux values and $\Bar{x}$ is the mean flux. Defined in this way, the PSD estimates have units of 1 per unit frequency (i.e. days). The power spectrum is computed in the rest frame of each source, that is, all $t_i$'s in the above equation are divided by $(1+z)$, including the time window, $\Delta t$. Periodograms are defined in the usual set of frequencies, $f_j=j/(N\Delta t)$, with $j=1,2,...,N/2$.

We could combine the periodograms from the seasonal segments of the ZTF light curves (there are typically five of them for each object), from the SDSS, the Gaia and the long-term binned ZTF light curves to construct a broad frequency band power spectrum for each quasar in our sample. However, periodogram estimates are distributed as $\chi^2$ variables with 2 degrees of freedom, and their error depends on the intrinsic power spectrum, which is unknown (e.g. \citealt{Papadakis1993}). Therefore, they are not suitable for fitting models to the PSD using $\chi^2$ techniques. For that reason, we followed the same approach as with P24, and we calculated ensemble power spectra for many quasars which have similar BH mass, luminosity, and redshift (as a proxy of rest-frame wavelength) as we explain below. 

\section{The broad-band power spectrum of quasars}
\label{sec:psd-shape}
The ensemble PSD analysis requires averaging the periodograms of quasars with similar physical properties, as outlined in P24. The availability of information on black hole mass, luminosity, redshift, and accretion rate (from the SDSS DR7 quasar catalogue by \cite{Shen2011}, see Sect. 2 of P24 for more details) allows us to explore how the PSD features (slope, normalisation, and bending/break frequencies) depend on these parameters. Compared to P24, the joint contribution to the PSD from SDSS, Gaia and long-term ZTF at low frequencies, and from the seasonal ZTF segments at high frequencies enable us to probe the quasar PSD over a wider frequency range, for any given bin of quasars with similar physical properties. P24 showed that quasar variability does not depend on redshift (at any given wavelength), but depends on \bhm, \lum\ (or \ledd=$L_{Bol}/L_{Edd}$), and rest-frame wavelength. However, the availability of simultaneous SDSS observations in \textit{ugriz} bands easily allowed P24 to select many bins of [\bhm, \ledd, z] and study how the PSD depends on them. Our sample of combined light curves has only the \textit{r} band, forcing us to select quasars in narrow redshift bins to not introduce a potential bias in the ensemble PSD due to mixing of sources with a significant large range of rest-frame wavelengths. This, together with the smaller sample of 4037 quasars, reduces our ability to study the PSD dependence on a broad range of BH masses and Eddington ratios because the number of adjacent [\bhm, \ledd, z] bins with enough sources to constrain their ensemble PSD accurately is more limited. Nonetheless, for each bin [\bhm, \ledd, z], we have a wider frequency range to constrain the PSD shape, while P24 had only three frequencies with $\Delta(\log \nu) \sim 0.5\ day^{-1}$.

\begin{figure}
    \includegraphics[width=1\columnwidth]{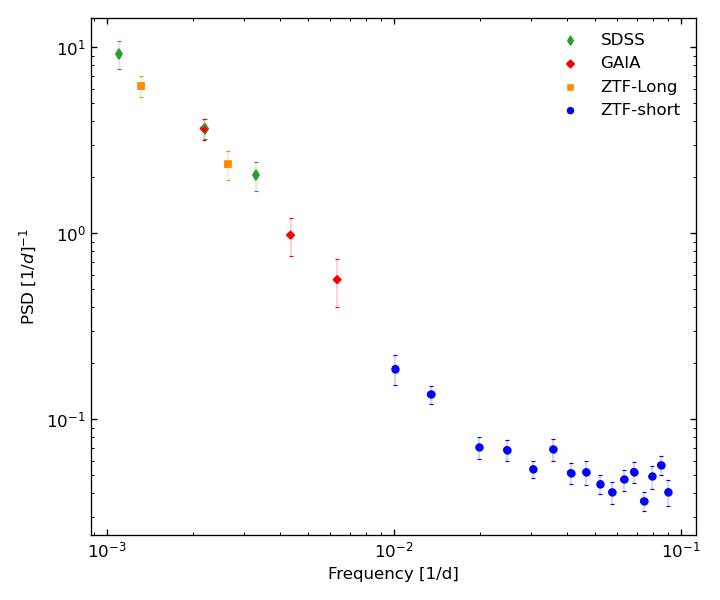}
    \caption{Ensemble power spectrum for the master sample of the quasars. 
Different colours and symbols show PSD estimates calculated using various light curves, as labelled (see Sect. \ref{sec:psd-shape} for details).}
    \label{fig:psd_combined}
\end{figure}

To probe the PSD shape, we initially focussed on a single subset of 69 quasars and the following physical parameters: 8.7 < \lbhm\ < 9.0, 45.9 < \llum\ < 46.1, and 1.25 < z < 1.45 (hereafter the master sample). This range of values corresponds to the peak of the distribution of the respective parameter values (see Fig.\,\ref{fig:Gaia_sdss_hist}); hence, the number of quasars is the highest in the narrowest bin and their ensemble power spectrum is ideal for studying its shape in detail. Given the mean redshift of the quasars in this sample, the observed light curves can probe intrinsic quasar variations in the UV band (i.e. around $\sim 2600$ \AA). Other sources of the larger sample are used in Sect. \ref{sec:trends} to discuss the dependence of the PSD on BH mass and accretion rate.

The combined PSD for the master sample is shown in Fig. \ref{fig:psd_combined}, with different colours and symbols for the different contributions. The points on the left side of the plot show the low-frequency PSD, calculated on timescales of the order of years, computed with the SDSS, Gaia, and long-term ZTF light curves. For each group of these light curves, we computed the 69 periodograms, we subtracted the Poisson noise, and calculated their mean and its error, in each frequency bin. For the SDSS and the long-term ZTF light curves the Poisson noise was directly measured by computing the variance of light curves from non-variable stars as a function of magnitude, as described in Appendix A of P24. For Gaia PSDs, where light curves from non-variable sources are not available, we computed the theoretical Poisson noise contribution as the mean error squared of the light curve points (which is representative of the light curve variance due to Poisson noise) divided by the frequency range over which we calculate the power spectrum and the light curve mean squared (since this is how we normalised the periodogram; see Eq.\,\ref{periodogram}).

On the right side of Fig.\,\ref{fig:psd_combined}, blue circles show the high frequency part of the ensemble PSD, calculated from the seasonal segments of the 12-day binned ZTF light curves. These points sample the power spectrum on time scales of weeks to months. To compute the ensemble PSD at these frequencies, we merged all the PSDs calculated from the approximately five seasonal PSDs of all the quasars in the master sample and computed the mean PSD (and its error) over bins of 100 points. The Poisson noise contribution was not subtracted from the high-frequency PSD estimates; instead, it was fitted a posteriori by adding a constant to the models we fit to the ensemble PSD.

The power spectrum estimates of the SDSS and the long-term ZTF light curves are consistent within the errors, as shown by the respective data points in the left part of Fig. \ref{fig:psd_combined}. This is also the case with the Gaia PSD estimates, despite the fact that Gaia measurements include contributions from a wide range of wavelengths (due to the width of its filter). According to P24, the UV/optical power spectra of the quasars depend on the wavelength, and this could affect the Gaia measurements, although Fig. \ref{fig:psd_combined} shows that this is not the case. The agreement of the ensemble power spectra estimated from light curves taken years apart and with different instruments implies that the rest-frame UV variability process remain constant over the timescales explored in this work ($\sim 20$ years). This is an important result and provides a strong indication that the long-term variability in AGN is stationary\footnote{Stationarity implies that the statistical properties of the process (like the mean, variance, auto-covariance function, and PSD) remain the same at all times.} on timescales of several years, although we cannot exclude non-stationarity at much longer timescales. The second result is that the variability amplitude continues to increase with increasing time scales, without clear evidence of a complete flattening of the PSD (to zero slope) down to about $10^{-3}$ days$^{-1}$ (i.e. $\sim 3$ years in rest-frame). 

\begin{figure}
    \centering
    \includegraphics[width=1\columnwidth]{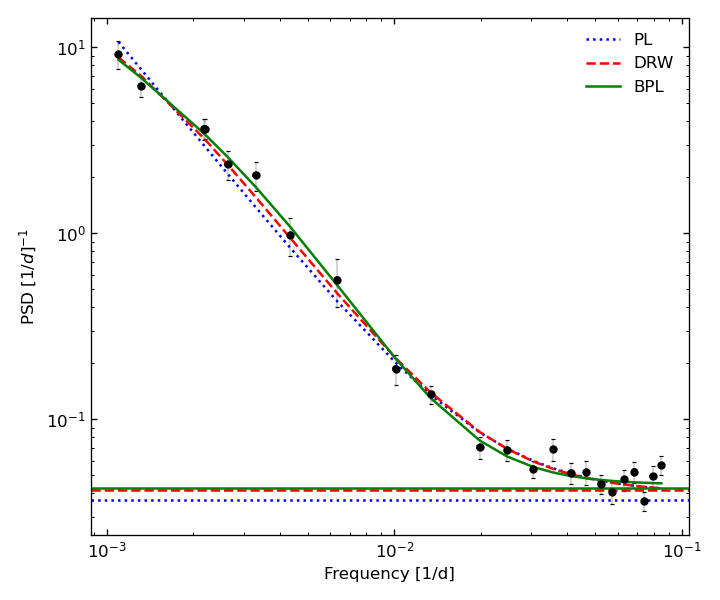}
    \includegraphics[width=1\columnwidth,height=0.7\columnwidth]{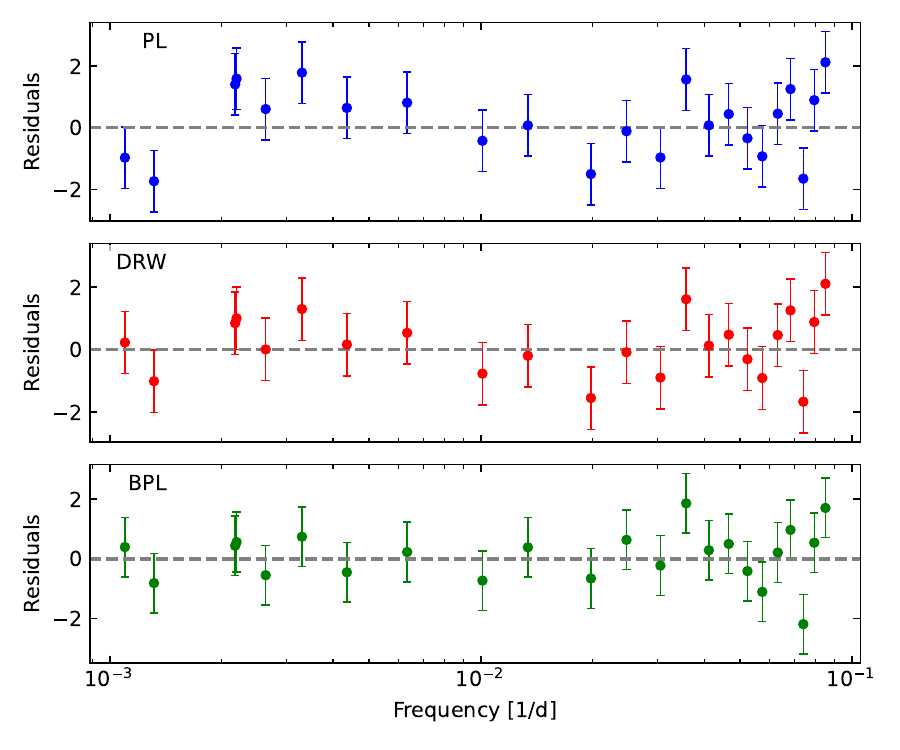} 
    \caption{{\it Top panel:}  Best-fit PL, DRW, and BPL models to the ensemble PSD of the master sample. Horizontal lines mark the noise floor level for each model. {\it Bottom panels:} Respective best-fit residuals plots; i.e. observed PSD--model+error.  }
    \label{fig:psd_combined_fit}
\end{figure}

\begin{table}
\caption{PL, DRW, and BPL best-fit parameters from the fits in Fig. \ref{fig:psd_combined}.}
\begin{center}
\begin{tabular}{lcccc}
\hline
\hline
Mod.  & $PSD_{norm}$\tablefootmark{a} & $\nu_{b}$\tablefootmark{b} & $\alpha$ & $\chi^{2}$/dof$\,(p_{null})$ \\
      &  & & & \\
PL & 3.4$\pm$0.2 & -- & --1.89$\pm$0.06 & 30.1/20\,(0.07)\\
DRW & 22$\pm$7 & $0.9\pm 0.2$ & -- & 22.2/20\,(0.32)\\
BPL & 0.011$\pm$0.002 & 3.5$\pm 1.1$ & --2.60$\pm$0.25 & 18.2/19\,(0.51)\\
\hline
\hline
\end{tabular}
\tablefoot{The parameter $\alpha$ represents the single slope of the PL model, and the high-frequency slope ($\alpha_{high}$) of the BPL model (where $\alpha_{low}$ is kept fixed at 1). DRW has no best-fit slope parameters, as $\alpha_{low}$ and $\alpha_{high}$ are fixed to 0 and 2, respectively. The break frequency, $\nu_b$ is not defined for the PL model.\\
\tablefoottext{a}{In units of day$^{-1}$,}
\tablefoottext{b}{In units of $10^{-3}$ day$^{-1}$. }
}
\end{center}
\label{tab:master}
\end{table}

To quantitatively determine the PSD shape, we fit the ensemble PSD of the master sample with a single power law, a bending power law, and the PSD associated with the DRW model, as we explain below. The single power law (PL) is defined as
\begin{equation}
    P(\nu) = PSD_{norm}\ (\nu/\nu_0)^{\alpha} + noise,
\label{eq:psd_single}
\end{equation}

\noindent where $\nu_0=0.002$ days$^{-1}$ is the frequency at which we evaluate PSD normalisation, $PSD_{norm}$. The more general bending power law (BPL) model takes the form
\begin{equation}
    P(\nu) = PSD_{norm} \nu^{\alpha_{low}}\bigg[1+\bigg(\frac{\nu}{\nu_b}\bigg)^{\alpha_{low}-\alpha_{high}}\bigg]^{-1} + noise.
\label{eq:psd_generic}
\end{equation}
\noindent This model was introduced by \cite{Mchardy2004}. It describes  the broad-band (i.e. low- and high-frequency) power spectrum of AGN in X-rays well. In the above equation, $\nu_b$ is the bending/break frequency above which the PSD slope steepens from $\alpha_{low}$ at $\nu<<\nu_b$, to $\alpha_{high}$ at $\nu>>\nu_b$. 
$PSD_{norm}$ is equal to $2PSD(\nu_b)\times \nu_b$, and equal to $PSD(\nu)\times \nu$ at frequencies $\nu << \nu_b$, for the BPL model with $\alpha_{low}=-1$. In this case, when the power spectrum is plotted in the $PSD(\nu) \times \nu$ representation, it appears flat below $\nu_b$. Its value is equal to $PSD_{norm}$, and for that reason, $PSD_{norm}$ is frequently called the `power spectrum amplitude', in the case of BPL models with $\alpha_{low}=-1$.
The power spectrum of the DRW model is equivalent to the BPL model with $\alpha_{low}=0$ and $\alpha_{high}=-2$. 

We fit the PL, BPL, and DRW models to the ensemble PSD of the master sample plotted in Fig.\,\ref{fig:psd_combined}. We kept $\alpha_{low}=-1$ fixed during the model fitting process of the BPL. If left free to vary during the fits, $\alpha_{low}$ is unconstrained, with best-fit values close to --1, but large errors due to the fact that the ensemble PSD does not extend to low enough frequencies. The best-fit results for the master sample are listed in Table \ref{tab:master}, while the best-fit models and residuals are plotted in Fig.\, \ref{fig:psd_combined_fit}.

All models provide statistically acceptable fits to the data, as $p_{null} > 0.01$ for all of them. However, the improvement of the BPL fit to the data is significant when compared with the PL best-fit. The difference in the best fit $\chi^2$ of the two models is $\Delta\chi^2$ = 11.9 for 1 extra degree of freedom. This is statistically significant according to the $F-$test (F statistic of 12.4, $p_{null}=0.002$). This result is strengthened and confirmed by the poor fits the PL model provides to the ensemble PSDs of quasars across various BH mass and accretion rate bins that are analysed in Sect. \ref{sec:trends}. We therefore conclude that the intrinsic PSD in the UV band of quasars with log(\bhm)\ $\sim 8.85$ and log(L$_{\rm bol})\sim 46$ shows a significant flattening at low frequencies, below $\nu_b\sim 1-3 \times 10^{-3}$ day$^{-1}$.

BPL provides the best fit to the PSD. The best-fit high-frequency slope is $-2.60\pm 0.25$ which implies a difference $\sim 2.4\sigma$ compared to the slope of $\sim -2$ predicted by the DRW model. However, the goodness-of-fit of the BPL model is not statistically significant when compared with the DRW best-fit to the ensemble PSD of the master sample. 

\begin{figure}
    \centering
    \includegraphics[width=1\columnwidth]{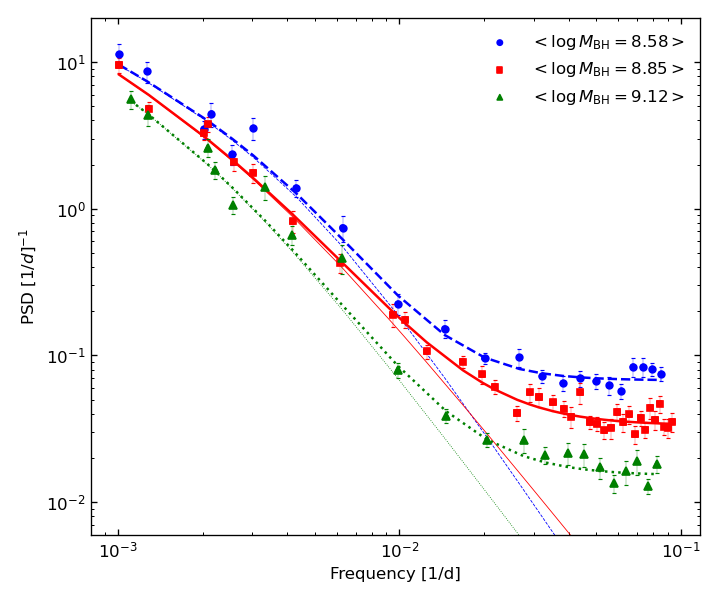}
    \includegraphics[width=1\columnwidth,height=0.7\columnwidth]{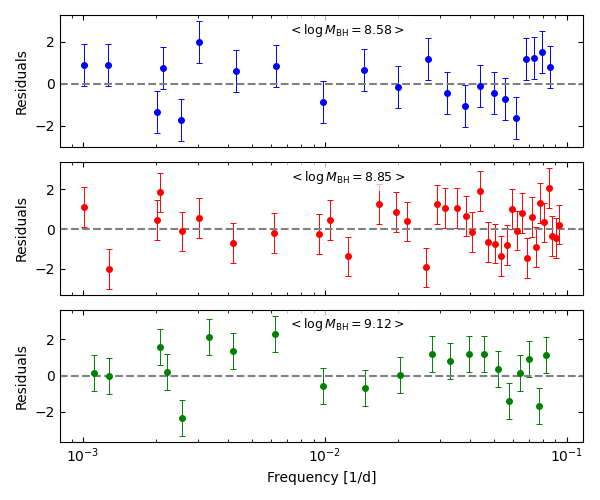} 
    \caption{{\it Top panel:} Ensemble power spectra for the quasars in the BHM1, BHM2, and BHM3 samples (mean values of the BH mass in each sample are reported in the legend). Dashed, solid, and dotted lines indicate the the best-fit BPL models to the power spectra. Thinner lines of the same colour and style are noise-subtracted. {\it Bottom panels:} Best-fit residuals. }
    \label{fig:psd_mass_fit}
\end{figure}

\begin{figure}
    \centering
    \includegraphics[width=1\columnwidth]{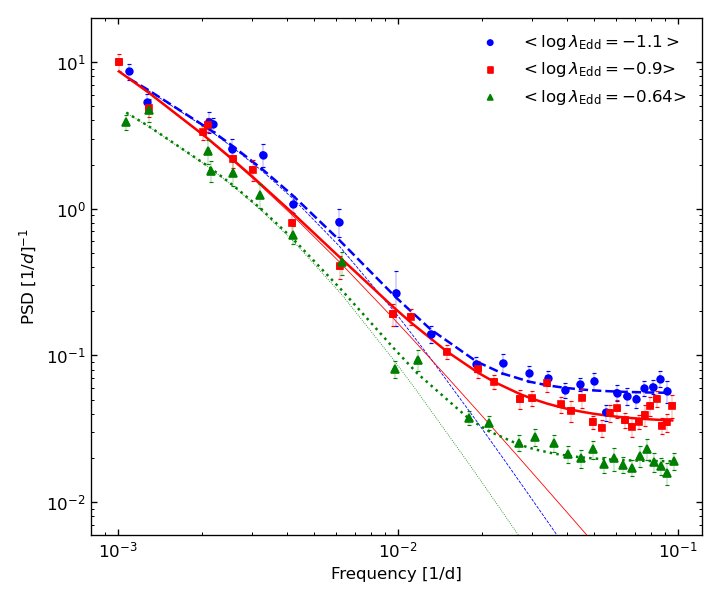}
    \includegraphics[width=1\columnwidth,height=0.7\columnwidth]{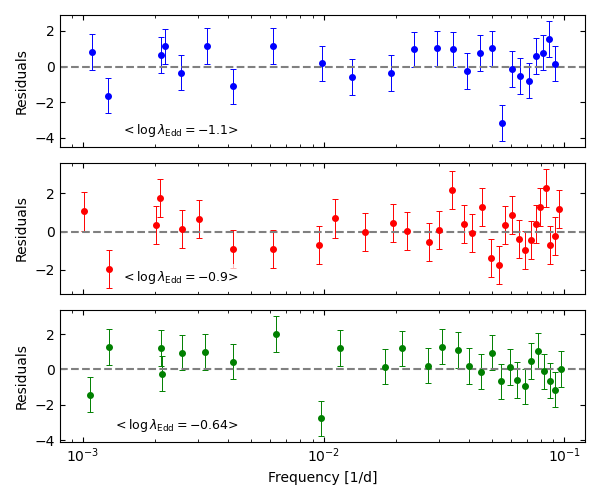} 
    \caption{Same as in Fig.\,\ref{fig:psd_mass_fit}, but for the AR1, AR2, and AR3 samples (mean values of the accretion rate for each sample are reported in the legend).}
    \label{fig:psd_ledd_fit}
\end{figure}

\section{The dependence of the quasar PSD on BH mass and accretion rate}
\label{sec:trends}

As in P24, we also investigated whether the power spectrum varies with \bhm\ and/or $\lambda_{Edd}$. To this end, we created six additional samples of quasars. Three of the samples include quasars with the same accretion rate (--1.0 < \lledd\ < --0.7) and redshift (1.15 < z < 1.45), but different BH mass masses, BHM1, BHM2, and BHM3, including quasars with 8.4 $\leq$ \lbhm\ < 8.7 (61 quasars), 8.7 $\leq$ \lbhm\ < 9.0 (113 quasars), and 9.0 $\leq$ \lbhm\ $\geq$ 9.30 (63 quasars), respectively. The other three samples include objects with the same BH mass (8.7 < \lbhm\ < 9.0) and redshift (1.15 < z < 1.45), but different accretion rates, AR1, AR2, and AR3, including quasars with 
--1.30 $\leq$ \lledd\ < --1.00 (69 quasars), --1.00 $\leq$ \lledd\ < --0.75 (96 quasars), and --0.75 $\leq$ \lledd\ $\geq$ --0.40 (66 quasars). We note that both the BHM2 and the AR2 samples are similar to the master sample in terms of the range of BH mass and accretion rates of quasars in them. We cannot calculate the ensemble PSD of quasars in additional bins to increase the range of the BH mass or the accretion rate values because the number of sources in each bin becomes too small and their distribution will not be Gaussian \citep{Papadakis1993}. We can increase the redshift bin size beyond 0.3, but this will result in considering quasars with rest-frame light curves in significantly different energy bands. Since the quasar variability depends on energy (see P24), the ensemble PSDs will be increasingly difficult to interpret. 

Figures \ref{fig:psd_mass_fit} and \ref{fig:psd_ledd_fit} show the ensemble power spectra (calculated as before for the master sample) for the quasars in the various BHM and AR samples. Clearly, the PSDs are not identical, which implies that their properties must vary with both the BH mass and the accretion rate. 

We fit the ensemble PSDs of the BHM and AR quasar samples with the DRW and BPL models. We did not consider the PL model, as it gives very poor fits to most of the data, which is in agreement with the result in the previous section. The best-fit results are listed in Tables \ref{tab:mass} and \ref{tab:ledd}. Statistically speaking, both the DRW and the BPL models fit the PSDs well (i.e. $p_{null}$ is larger than 0.01 in all cases). The only exception is the BHM3 PSD, where $p_{null}$ is low for both best-fit models. The respective residuals plot in Fig.\,\ref{fig:psd_mass_fit} shows that the BPL model fits well the overall PSD shape, but the PSD is rather noisy, hence, the low $p_{nul}$.  

BPL provides a better fit to all ensemble power spectra, for both the BHM and AR samples. Based on the assumption that there is an underlying universal model that has to fit the PSDs in all bins, the total $\chi^2$ for the best-model fits in the case of the DRW model is 183.8 for 129 degrees of freedom. The respective numbers are $\chi^2_{total}=152.2$ for 124 degrees of freedom in the case of the BPL model (excluding the best-fit models to the BHM3 PSD for both models). The goodness of the DRW fit to all PSDs is poor ($p_{null}=0.0012$), while the BPL provides an acceptable fit to all PSDs ($p_{null}=0.02)$. Furthermore, the improvement in the goodness of fit in the case of the BPL model ($\Delta\chi^2=24.5/5$ dof) is statistically significant according to the F-test (F-statistic $3.8$, $p_{null}=0.0029)$. We therefore conclude that, in general, the quasar power spectra are not consistent with the predictions of the DRW model.

The ensemble PSDs plotted in Figs.\,\ref{fig:psd_mass_fit} and \ref{fig:psd_ledd_fit} can be used to investigate the dependence of the PSD characteristics (i.e $\alpha_{high}$, $PSD_{norm}$ and $\nu_b$) on the physical parameters of the quasars, that is, \bhm\ and \ledd. We find that $\alpha_{high}$ does not depend on either \bhm\ or \ledd. All the best-fit $\alpha_{high}$ values listed in Tables \ref{tab:mass} and \ref{tab:ledd} remain consistent with the high-frequency slope of the ensemble PSD listed in Table \ref{tab:master}. The weighted mean of $\alpha_{high}$ in the master, BHM1, BHM3, AR1, and AR3 samples (which include objects which do not appear in the other samples) is $-2.7 \pm 0.1$ and this is our best estimate of the high-frequency slope in quasars (in the UV band). 

All the best-fit values of $PSD_{norm}$ listed in Table \ref{tab:mass} are consistent with the best-fit $PSD_{norm}$ of the master sample, implying that $PSD_{norm}$ does not depend on \bhm. However, the best fit value of $PSD_{norm}$ for the AR3 sample [log(\ledd$)\sim-0.64$] is approximately half that of $PSD_{norm}$ in the other two AR samples listed in Table \ref{tab:ledd}, and the difference is highly significant. These results indicate a possible decrease in PSD amplitude at accretion rates greater than $10^{-0.64}$ (i.e. $\sim 0.23$ of the Eddington limit). 

Finally, we also considered the dependence of the bending frequency, $\nu_b$, on the BH mass and accretion rate. Regarding the latter parameter, we did not notice any trend between $\nu_b$ and \ledd. The best-fit values $\nu_b$ in Table \ref{tab:ledd} are consistent (within the errors) with the best-fit bending frequency in the PSD of the master sample. However, we noticed a trend of decreasing bending frequency with increasing BH mass, in Table \ref{tab:mass}. A straight line (in the log--log space) with a slope of $-0.6\pm 0.1$, fits the $\nu_b$ vs \bhm\ plot well. This implies an anti-correlation between the bending frequency and BH mass, although it is based on only three points. 

However, if $PSD_{norm}$ and $\alpha_{high}$ are not dependent on \bhm, then the relation $\nu_b\propto$ \bhm$^{-0.6}$ we find can explain the difference between the PSDs shown in Fig. \ref{fig:psd_mass_fit}. The PSD amplitude at low frequencies [in PSD$(\nu)\times\nu$] could be the same in all objects, but because the bending frequency decreases with increasing BH mass, the PSD amplitude in the sampled frequency range appears to decrease with increasing BH mass. In the case of the PSDs in Fig. \ref{fig:psd_ledd_fit}, the main difference is between the \ledd\ $\approx 0.23$ PSD (green triangles) and the other two PSDs. The lower amplitude of this PSD could, in fact, be due to a difference in $PSD_{norm}$, which can decrease with increasing accretion rate. The ratio between the normalisation in the first and last bin is $PSD_{norm}$(AR3)/$PSD_{norm}$(AR1) $\sim 0.6\pm0.1$.

\begin{table}
\caption{Best-fit parameters when fitting the ensemble PSDs in various \bhm\ bins, for quasars with \lledd\ in the range [-1.0--0.7].}
\begin{center}
\begin{tabular}{lllllll}
\hline
\hline
 & BHM1 & BHM2 & BHM3  \\
$\log{M_{BH}\in}$ & \small{[8.4 , 8.7)} & \small{[8.7 , 9.0)} & \small{[9.0 , 9.3]}  \\
\hline
\small{num. sources} & 61 & 113 & 63 \\
$PSD_{norm,\ DRW}$\tablefootmark{a} & 29$\pm$12 & 18$\pm$4 & > 8  \\
$PSD_{norm,\ BPL}$\tablefootmark{b} & 1.0$\pm$0.2 & 1.0$\pm$0.2 & $ 0.8\pm$0.2  \\
$\nu_{b,\ DRW}$\tablefootmark{c} & 0.8$\pm$0.2 & 0.9$\pm$0.1 & < 5  \\
$\nu_{b,\ BPL}$\tablefootmark{c} & 4$\pm$1 & 2.7$\pm$0.9 & 2.1$\pm$0.8  \\
$\alpha_{,\ DRW}$ & -- & -- & --  \\
$\alpha_{high,\ BPL}$ & --2.8$\pm$0.3 & --2.4$\pm$0.2 & --2.5$\pm$0.2  \\
$\chi^{2}/dof_{,\ DRW}$  & 32.6/19 & 42.9/34 & 44.3/18  \\
$\chi^{2}/dof_{,\ BPL}$  & 25.2/18 & 40.7/33 & 32.8/17  \\
$p_{null,\ DRW}$ & 0.026 & 0.14 & 0.0005 \\
$p_{null,\ BPL}$ & 0.12 & 0.17 & 0.001\\
\hline
\hline
\end{tabular}
\end{center}
\tablefoot{
\tablefoottext{a}{In units of (1/day$^{-1}$),}
\tablefoottext{b}{In units of $10^{-2}$(1/day$^{-1}$),}
\tablefoottext{c}{In units of $10^{-3}$ day$^{-1}$.}
}

\label{tab:mass}
\end{table}

\begin{table}
\caption{Best-fit parameters when fitting the ensemble PSDs in various $\lambda_{Edd}$ bins, for quasars with log(\bhm) in the range [8.7--8.9].}
\begin{center}
\begin{tabular}{lllllll}
\hline
\hline
 & AR1 & AR2 & AR3  \\ 
$\log{\lambda_{Edd}}\in$ & \small{[--1.3 , --1.0)} & \small{[--1.0 , --0.75)} & \small{[--0.75 , --0.40]}  \\\hline\smallskip
\small{num. sources} & 69 & 96 & 66 \\
$PSD_{norm,\ DRW}$\tablefootmark{a} & 14$\pm$3 & 16$\pm$4 & 10.3$\pm$2.5  \\
$PSD_{norm,\ BPL}$\tablefootmark{a} & 9$\pm$1 & 12.0$\pm$0.3 & 5.20$\pm$0.06 \\
$\nu_{b,\ DRW}$\tablefootmark{b} & 1.2$\pm$0.2 & 1$\pm$0.1 & 0.9$\pm$0.1  \\
$\nu_{b,\ BPL}$\tablefootmark{b} & 5$\pm$1 & $<5$ & 4.1$\pm$0.8  \\

$\alpha_{,\ DRW}$ & -- & -- & --  \\
$\alpha_{high,\ BPL}$ & --2.82.8$\pm$0.3 & --2.2$\pm$0.2 & --2.8$\pm$0.2  \\

$\chi^{2}/dof_{,\ DRW}$  & 33.6/22 & 34.2/29 & 40.5/25  \\
$\chi^{2}/dof_{,\ BPL}$  & 28.4/21 & 34.4/28 & 30.5/24  \\

$p_{null,\ DRW}$ & 0.05 & 0.23 & 0.03 \\
$p_{null,\ BPL}$ & 0.13 & 0.19 & 0.17 \\
\hline
\hline
\end{tabular}
\end{center}
\tablefoot{
\tablefoottext{a}{In units of $10^{-3}$(1/day$^{-1}$),}
\tablefoottext{b}{In units of $10^{-3}$ day$^{-1}$.}
}

\label{tab:ledd}
\end{table}


\section{Discussion and conclusions}
\label{sec:discussion}
We computed the ensemble power spectrum of hundreds of quasars, which are part of the SDSS Stripe-82 sample of P24, in the rest-frame UV band and in various BH mass and accretion rate bins. We were able to estimate the power spectrum over two orders of magnitude in frequency, from $\sim 1/(10$ days), to $\sim 1/(1000$ days). This is a significant advance compared to our previous work (P24), where we were able to estimate the quasar power spectra in just three frequencies, over a narrower range. The advance was made possible because we considered data from two additional surveys in addition to SDSS (ZTF and Gaia), allowing us to extend the PSD at high frequencies and better constrain it at low frequencies. We estimated the power spectrum using traditional Fourier techniques. As a result, the PSD units are well defined, and the UV PSDs can be directly compared with AGN PSDs estimated in other energy bands (i.e. X-rays). However, cross-matching data from multiple surveys resulted in limiting the original P24 sample to half of the sources and a single photometric band (to limit cross-calibration biases in combining the light curves; see Appendix \ref{subsec:agn_lc_calib}). As UV/optical variability of AGN depends on energy (e.g. P24), we restricted the analysis to a narrow redshift range $1.25 < z < 1.45$, resulting in PSDs at $\sim 2600$\AA\ rest-frame (i.e. the UV band). While P24 was able to exploit a larger sample and five filters \textit{ugriz} to get many [\bhm, \ledd] bins at any given wavelength, this work focuses on a limited number of bins spanning $\lesssim 1$ dex in each parameter. The number of sources in other bins was significantly smaller than the minimum necessary for the statistical properties of the ensemble PSD estimates to be appropriate for model fitting. Nonetheless, for each bin we have a wider frequency range to better constraint the intrinsic PSD shape. Our main results are summarised below.
\begin{enumerate}
    \item We were able to accurately estimate the ensemble PSD of quasars in three BH mass bins with an average \bhm\ = $3.8\times 10^8, 7.1\times 10^8,$ and $1.3\times 10^9$ M$_{\odot}$, and with three accretion rate bins with an average \ledd\ = 0.08, 0.13, and 0.23. 
    \item We find that a power-law model does not fit the PSDs well. We detected bending frequencies, $\nu_b$, in the ensemble PSDs in all BH mass and accretion rate bins. This is the first time that bending frequencies have been detected directly in the UV power spectra of a single sample of luminous quasars, without the need to combine PSDs calculated on narrower frequency ranges and then rescaling frequencies according to pre-defined relations between $\nu_b$ and BH mass. 
    \item A bending power-law model fits all PSDs well, with a mean slope at high frequencies, above $\nu_b$, of $-2.7\pm0.1$ depending neither on \bhm\ nor on \ledd. The slope at low frequencies is consistent with --1 in all ensemble PSDs (irrespective of the BH mass and/or accretion rate).
    \item The DRW model does not fit the ensemble PSDs equally well. A low frequency slope of zero (as expected in the case of the DRW model) is not consistent with the data (as the best-fit results worsen considerably, when compared with the BPL best-fit models). Furthermore, the high-frequency BPL best-fit slope of $-2.7\pm0.1$ is significantly different from the DRW prediction of $-2$. 
    \item The amplitude of the BPL model, $PSD_{norm}$, does not depend on \bhm, although it might decrease with increasing accretion rate. The ratio between the normalisation in the first and last bin, $PSD_{norm}$(AR3)/$PSD_{norm}$(AR1), is $\sim 0.6\pm0.1$. The possible existence of a trend is tentative and we need to study ensemble PSDs of quasars on a wider range of accretion rates to reach conclusive results.
    \item The bending frequency decreases with BH mass, approximately as $\nu_b \propto M_{BH}^{-0.6\pm0.1}$. We did not detect any dependence of $\nu_b$ on the accretion rate but, as in the case of $PSD_{norm}$, we need to fit the ensemble PSDs of quasars over a wider range of accretion rates to reach conclusive results.
\end{enumerate}

\begin{figure}
    \centering
    \includegraphics[width=1
    \linewidth]{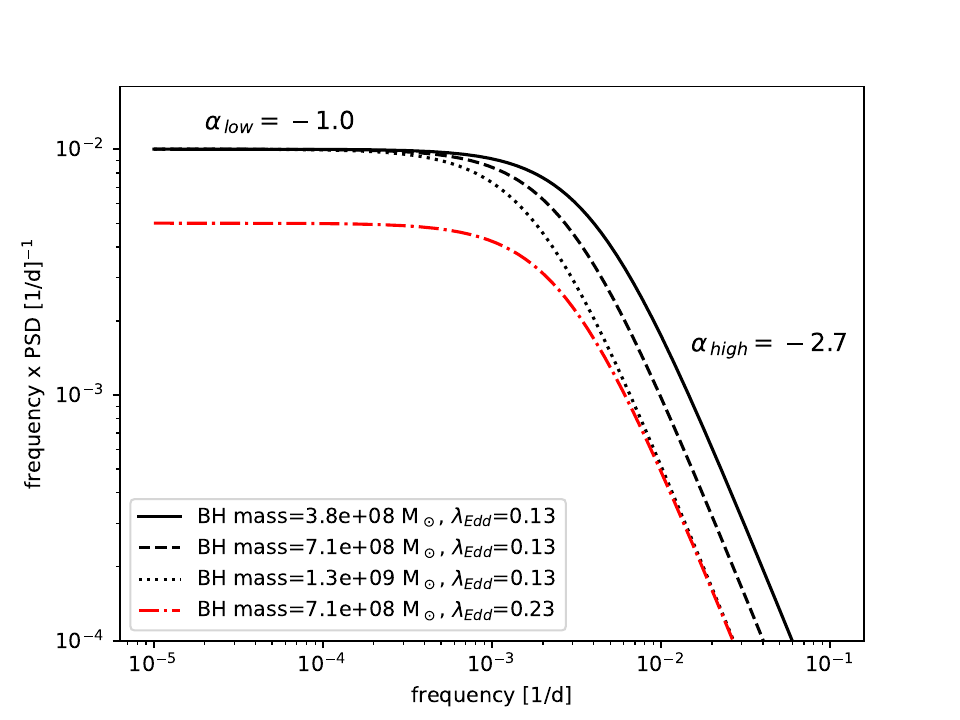}
    \caption{Schematic representation of the best-fit bending power-law PSD model and its scaling with BH mass and accretion rate.}
    \label{fig:PSDmodel}
\end{figure}

Figure \ref{fig:PSDmodel} shows a schematic representation of the quasar's power spectrum in the UV band, based on our results. The power spectrum in this figure is plotted in the $PSD(\nu)\times \nu$ representation. The low-frequency PSD slope was fixed at --1. If we leave $\alpha_{low}$ free to vary during model fitting, the best-fit results are still consistent with --1, but the errors are large (not only in $\alpha_{low}$ but also in the other parameters). The dependence of the bending frequency on BH mass, as well as the possible dependence of the PSD amplitude on the accretion rate, are based on the results of fitting the ensemble PSDs over three bins covering less than 1 dex in BH mass and \ledd. Clearly, it is important to extend the study of the ensemble power spectrum of quasars to many more BH mass and accretion rate bins to better constraints the scaling relations. An important factor that determines the size of the sample and the number of quasars in narrow BH mass and accretion rate bins is the use of the Gaia light curves. The future Gaia data release will hopefully include more quasars and additional epochs overlapping with the beginning of ZTF. This, together with efforts in populating the gaps between the surveys and/or investigating sophisticated interpolation methods, could be used to (reliably) produce evenly spaced time series using as input light curves such as the one plotted in Fig.\,\ref{fig:agn_lc_comb} and to study the PSD on an even larger frequency range. We discuss this topic further in this section. In the future, we also plan to explore other methods to estimate the ensemble power of quasars, which will require fewer objects per bin. Despite the cautious remarks above, Fig. \ref{fig:PSDmodel} could be used to gain important information on the UV variability of quasars. 

\subsection{Comparison with X-rays and additional physical constraints}

The best-fit PSD normalisation when we fit the ensemble PSD of the master sample with the BPL model is 0.011$\pm 0.002$ (day)$^{-1}$. Since $a_{low}=-1$ in Eq.\,\ref{eq:psd_generic}, then PSD$(\nu)\times \nu$ will be equal to $0.011$ at very low frequencies (i.e. when $\nu<<\nu_b$). Interestingly, this is also the case with the X-ray PSDs of AGNs. The results indicate that the X-ray PSD normalisation is roughly the same in all AGNs we have studied so far, and that PSD$_{\rm X-rays}(\nu)\times \nu\sim 0.01-0.02$ at frequencies much lower than the break frequency. For example, \cite{Paolillo23} assumed a BPL model (with $a_{low}=-1$) for the X-ray PSD and found that A=0.016$\pm0.003$ Hz$^{-1}$ for a large sample of AGNs, which also includes very bright quasars. This implies that  PSD$_{\rm X-rays}(\nu)\times\nu\sim 0.016$ at frequencies lower than $\nu_b$. Furthermore, \cite{papadakis24} studied the 14-195 keV band power spectrum of the brightest AGN in the BAT survey sample. They found that the PSD is consistent with a power-law model with a slope of $-1$ at low frequencies (i.e. at frequencies lower than the expected break frequency) and they also found that $PSD_{\rm X-ray}(\nu)\times \nu=0.014\pm 0.003$ at these low frequencies. These results indicate that when normalised to the (square) of the mean flux, the PSD amplitude at frequencies much lower than the break frequency (i.e PSD$(\nu)\times \nu$) is $\sim 0.01-0.015$, both in X-rays and in the UV band. This is an interesting result. It may be a coincidence, but it is also  consistent with X-ray reverberation. \cite{Papoutsis25} showed that the observed UV/optical variability in SDSS quasars, from 1300\AA\ to 4000\AA, could be the result of X-ray reverberation, if the X-ray corona is powered by the accretion power, the black hole spin is lower than $\sim 0.7$, and the height of the X-ray corona is in the range of 20-40$R_g$, where $R_g=G$\bhm$/c^2$ is the gravitational radius of the BH.

The best-fit bending frequencies in the UV power spectra can also be used to constrain various models. In the case of X-ray reverberation, we expect two breaks in the UV (and optical) PSD \citep[see][]{Papoutsis25}. The first is the bending frequency that also appears in the X-ray PSD. However, the bending timescale, $T_{b,UV}=1/\nu_b$, is $\sim 285$ days for the master sample (see Table \ref{tab:master}). The respective timescale in X-rays for a quasar with \bhm\ $ \sim 7\times 10^8$M$_{\odot}$ and \ledd\ $ \sim 0.13$, is $T_{b,X-rays}\sim 22$ days, assuming the \cite{mchardy06} relation. Therefore, the bending frequencies we detect cannot be associated with the bending frequency in the X-ray PSD.

The second bending frequency in the UV/optical PSDs should be due to the break frequency in the disc transfer function. Roughly speaking, this bending time scale is associated with the time it takes for the X-rays to propagate from the inner disc up to the outer radius of the part of the disc that emits in the UV. Interestingly, \cite{Panagiotou22} predicted that this PSD break should be proportional to \bhm$^{-0.65}$ (see their Eq. A2), which is very similar to what we found. This result supports the X-ray reverberation hypothesis.

However, the bending timescales that we detect in the UV PSDs are rather low. For example, \cite{Papoutsis25} showed that the bending frequency in the disc transfer function should be around $\sim 10^{-2}$ day$^{-1}$ for an AGN with a mass of $8\times 10^8$ and an accretion rate of 0.1 of the Eddington limit (see Fig. 1 in their paper). However, the best fit $\nu_b$ in the combined PSD of the master sample is about two to three times smaller. We need PSDs for many more objects, preferably sampled at even lower frequencies, to perform tests with theoretical models.

The other possibility is that the observed optical/UV variations are due to intrinsic disc variations. \cite{Liubarskii97} proposed a detailed model with respect to the variability of the flux of the disc. They showed that the resulting PSD will have a slope of --1 if there are fluctuations of the accretion rate that propagate inward, and their characteristic timescale is on the order of the local viscous timescale. In this case, the bending timescale should correspond to the viscous time at the inner disc radius. If that is equal to 6$R_g$ (in the case of BH with spin zero), then the expected viscous time scale for a $\sim 7\times10^8$ M$_{\odot}$ is $\sim 600$ days, assuming that the radius vs disc height ratio is 10, and the viscosity parameter is 0.1 \citep{Paolillo2025}. This is twice the timescales we observed in our work. Perhaps a different viscosity parameter and/or disc radius-to-height ratio could explain the observed discrepancy. Another possibility could be that the flux of the disc varies on the thermal time scale. For the PSD of the master sample (with \bhm$\sim 7\times 10^8$M$_{\odot}$), 
the break time scale of $\sim 300$ days in the PSD could be equal to the thermal time scale at $\sim 80R_g$ (assuming that the viscosity parameter is 0.1). It is not clear why the disc should be variable on the thermal time scale at this radius. In any case, a model should not only match the observed time scale with some theoretical timescale, but should also be able to explain the amplitude and slope of the PSD both at low and high frequencies. 

More recently, \cite{Sun2020} proposed a Corona-Heated Accretion-disk Reprocessing (CHAR) model, which fits the data with steeper than DRW slopes at high-frequencies, and no strong evidence for a complete flattening with light curves over than 20-years long (\citealt{Zhou2024}). Another possibility is that variability at low and high frequencies are associated with two different components, accretion rate fluctuations and X-ray reprocessing, respectively (e.g. \citealt{Yu2025}).


\subsection{Comparison with P24 and other past results}

P24 estimated the ensemble PSD of SDSS quasars over many bins of BH mass, accretion rate and rest-frame wavelength, collectively covering a frequency range from 250 to 1500 days (rest-frame). However, as the 
SDSS light curves used for the analysis were six years long, they were limited to three frequencies per bin and unable to directly detect any bending time scale in the power spectrum. They found that the PSD over the SDSS range could be fitted well with a PL model and that the amplitude and slope of the best-fit PL models depend on \bhm\ and \ledd, as follows: $\log(PSD_{\rm amp,P24}) \propto -0.5\log($\bhm$)-0.7\log($\ledd$)$, and $PSD_{slope,P24}\propto -0.5\log($\bhm$)-0.4\log($\ledd$)$. Our results, taking advantage of the additional frequencies, put better constraints on the intrinsic PSD shape and can be used to understand the P24 relations. 

We plotted the P24 PSD parameters (i.e. $\log(PSD_{\rm amp,P24})$ and $PSD_{slope,P24}$ with the best-fit PSD parameters we determined (i.e. $\nu_b,$ and $PSD_{norm,BPL}$). We find that $PSD_{amp,P24}\propto\nu_b^{0.7}$ and $PSD_{slope,P24}\propto 0.7\log(\nu_b)$. Since $\nu_b \propto$ \bhm$^{-0.6\pm 0.1}$ (see Sect. \ref{sec:trends}), then we can easily explain the dependence of $PSD_{amp,P24}$ and $PSD_{slope,P24}$ on BH mass as the result of $\nu_b$ shifting across the narrow frequency range sampled by P24 (as also discussed in their Fig. 19). 
The dependence of $PSD_{amp,P24}$ and $PSD_{slope,P24}$ on the accretion rate could be due to a possible dependence of $\nu_b$ on \ledd. As discussed in Sect. \ref{sec:trends}, we do not detect a correlation between $\nu_b$ and \ledd\,, but this may be due to the limited range and number of accretion rate bins in which we could estimate ensemble PSDs. It is also possible that the dependence of $PSD_{amp,P24}$ and $PSD_{slope}$ on the accretion rate may be due to the decrease of $PSD_{norm}$ with increasing \ledd. However, we cannot confirm a correlation between $PSD_{amp,P24}$ and $PSD_{slope,P24}$ with $PSD_{norm,BPL}$ in the limited range of \ledd\ we considered in this work. 

The relation between break timescales and BH mass is found in many recent works, using independent methods and datasets (e.g. \citealt{Tang2023}; \citealt{Arevalo2024}; \citealt{Yuk2025}), although there is no clear consensus on the exact dependence as the comparison between explicit PSD calculation, CARMA model fitting or SF analysis is not trivial. \cite{Burke+21} fit the light curves of $\sim 70$ quasars with a DRW model, and found that the characteristic bending timescale is related to the BH mass as $\tau \propto$ \bhm$^{0.4\pm0.05}$. This relation is flatter than the one we determined here ($\nu_b \propto M_{BH}^{-0.6\pm0.1}$). Furthermore, we used their Eq. (1) and computed the expected $\nu_b$ for quasars with the mean BH mass in the BHM1, BHM2, and BHM3 samples. We found that the \cite{Burke+21} predictions for $\nu_b$ are systematically $\sim 1.6$ times smaller than our best-fit values. The (significant) discrepancy between our results and those of \cite{Burke+21} is probably due to the fact that these authors assumed that the light curves are well fitted by a DRW model and computed the associated timescale, while we fit the ensemble PSD with a general BPL.

A work similar to P24 was also presented by \cite{Arevalo2024}, who used $g-$band light curves obtained by ZTF through their public data release DR14, and considered a large number of quasars in the small redshift range of $0.6 < z < 0.7$, over bins of BH mass and accretion rate. Their redshifts imply a rest-frame wavelength of $\sim 2900$\AA, which is close to the one presented in this work. They initially assumed the \cite{Burke+21} correlation between timescale and BH mass to shift the frequency of individual PSDs over a broader frequency range. Then, they fit all PSDs with BPL models at various fixed slopes, allowing for a dependence on both BH mass and accretion rate. The final best-fit has low- and high-frequency slopes of --1 and --3, respectively, with $\nu_b \propto M_{BH}^{-0.67\ to\ -0.5} \times \lambda_{Edd}^{-0.41\ to\ -0.12}$. While we are in agreement with the dependence of $\nu_b$ on BH mass, we do not find and trend with the accretion rate and our mean high-frequency slope of --2.7 is a bit flatter (although --2.5 is their second-best fit, and their procedure is fixing the slope a priori while to us is a free parameter). We also find lower PSD amplitudes, which may be due to the different technique to calculate the PSD, and a stronger dependence of $PSD_{norm}$ on the accretion rate [$PSD_{norm}$(AR3)/$PSD_{norm}$(AR1) $\sim 0.6\pm0.1$, while according to \cite{Arevalo2024} it should be $\sim 0.3$, using their Eqs. 3 and 4, the values listed in their Table 3, and the mean accretion rate and the BH mass of the quasars in the AR3 and AR1 bins]. 

We believe the main reason for the differences between the \cite{Arevalo2024} results and our work stems from the use of the model-dependent assumption adopted to shift PSDs. This clearly shows that it is important to estimate the power spectra of quasars over the broadest possible frequency range to detect the bending frequency and the low-frequency slope with high accuracy. At the same time, a proper characterisation of the PSD dependence on the AGN physical properties requires having a large sample of sources to be binned in many subsets. While analyses such as P24 and \cite{Arevalo2024} definitely satisfied the second requirement, the present work is aimed at constraining the PSD shape of fewer quasar samples with the most model-independent approach as possible. As the study of AGN variability becomes more and more detailed, the general agreement is that DRW can be used as a first order approximation, but it does not  accurately describe the results. Several works report of high-frequency PSD slopes steeper than the --2 value of the DRW, consistent with our findings, along with possible evidences of multiple break frequencies (e.g. \citealt{Mushotzky+11}; \citealt{Tachibana2020}; \citealt{Stone2022}). Analogous conclusions have  also been reported by \cite{Yu2025}, who analysed a sample of combined AGN light curves similar to ours, showing that a damped harmonic oscillator (i.e. a higher order CARMA model) fits the data significantly better than a DRW.


\subsection{Very-low-frequency PSD from combined light curves: Attempts at interpolation}
\label{sec:comb_int}
The only way to determine the PSD at frequencies lower than the the range of this work is to combine light curves from as many different surveys as possible and compute a unique PSD from them. This requires a considerable effort in calibration and homogenisation, but also accounting for the gaps between the surveys with no overlapping region. In the light curves we initially considered, albeit photometry was cross-calibrated (Appendix \ref{subsec:agn_lc_calib}), the extent of such gaps and their number prevent us from using a simple linear interpolation to account for the missing points, as we did for the high-frequency ZTF data. Several advanced techniques have been proposed to deal with non-uniform time series (e.g. \citealt{Lefkir2025}), but the biggest challenge is to avoid (or limit) the addition of spurious variability and keep the analysis model-independent. Moreover, when combining data from different surveys, ensuring there are no residual offsets between them is fundamental and requires applying the same approach on non-variable objects (which we could not do with Gaia DR3).  

A possibility is to model the light curve with a continuous function based on the measured points, and then use yearly-spaced epochs to measure the PSD. We tested a Radial Basis Function (RBF) interpolation method, which is commonly used to get a continuous approximation of a dataset from scattered and irregular points, such as uneven time series. More details about the implementation, the choice of parameters and its limitations are provided in the Appendix \ref{sec:rbf_int}. We focused on the master sample of 69 quasars defined in Sect. \ref{sec:psd-shape} and used the full r-band light curves, including the early SDSS epochs and the PS1 data, as in Fig. \ref{fig:agn_lc_comb}. The RBF-interpolation procedure produced evenly sampled light curves with a 356-day cadence, which we used to compute the PSD as described in Sect. \ref{sec:PSD_estimate}. While at high frequencies the reconstructed PSD-shape from the interpolation is strongly dependent on the input parameters (i.e. kernel function, smoothing, polynomial degree, sampling cadence), at frequencies $\nu \lesssim 2 \times 10^{-3}\ day^{-1}$ the outcomes are more robust for any reasonable input combination. 

\begin{figure}
    \includegraphics[width=1\columnwidth]{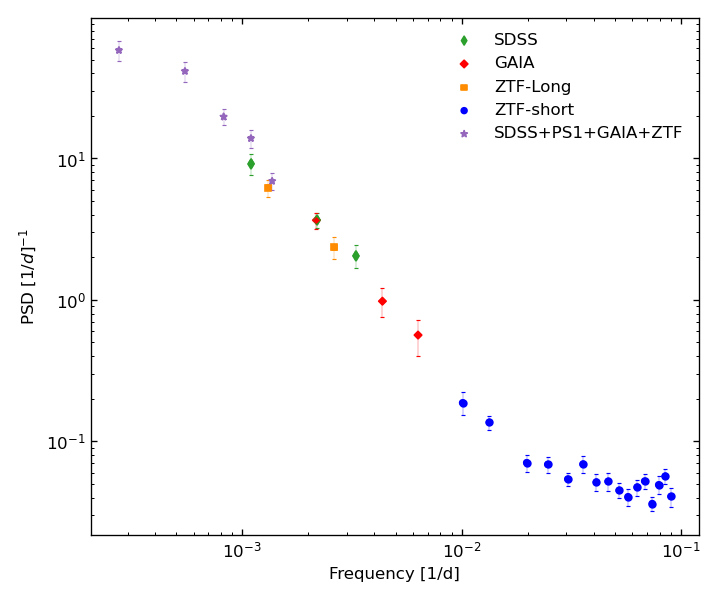}
    \caption{Ensemble power spectrum for the master sample of quasars, as in Fig. \ref{fig:psd_combined}, with the addition of estimates from the RBF-interpolated light curve combining all the surveys (purple stars).}
    \label{fig:psd_combined_int}
\end{figure}

The results of the interpolated PSD are shown in Fig. \ref{fig:psd_combined_int}, as purple points on top of the other independent estimates from the single surveys. There is a good agreement in the overlapping region, of around $\nu \sim 10^{-3}\ day^{-1}$. The key improvement is that the new light curves allow for the calculation of the PSD at frequencies that are about ten times lower than the lowest frequency in the PSDs we studied. The PSD at these frequencies appears to still increase, but at a lower rate. If we fit the new PSD with the same BPL model as before, we found best-fit parameters that are consistent within the errors with those of Fig. \ref{fig:psd_combined_fit}. However, the new, low-frequency points appear to have a higher normalisation than the other PSD points. This is a potential issue, as it is possible that creating an evenly sampled light curve via RBF interpolation could introduce additional variance to the light curves. To have a fully reliable interpolation method, thorough simulations of all possible outcomes are required. The tests we performed on simulated DRW light curves show that the method is potentially able to reconstruct the mean long-term variability trends, but the reliability of the reconstructed ensemble PSD depends on the position of the break and the variability amplitude (see Appendix \ref{sec:rbf_int}). This is a critical point for all the possible light curve fitting or interpolation techniques, which can potentially bias the results of the analysis. Given these issues, we did not measure the ensemble PSD on the combined light curves, but we did choose to present  our preliminary attempts to create evenly sampled light curves, combining data from various surveys as they might be useful for future works. In fact, this will be the only way to probe the quasar long-term variability, until LSST survey light curves are delivered a few years from now. However, even following the start of LSST, such approaches will be crucial to investigating AGN variability on timescales of over a decade, as well as for high-redshift samples where the probed rest-frame timescales are much shorter than those of local AGNs.

\begin{acknowledgements}
MP, VP, DD and MF acknowledge the financial contribution from PRIN-MIUR 2022 and from the Timedomes grant within the ``INAF 2023 Finanziamento della Ricerca Fondamentale''. IEP acknowledges support from the Visiting Professor programme of the Federico II University. DD also acknowledges PON R\&I 2021, CUP E65F21002880003. CMR and MIC acknowledge financial support from the INAF Fundamental Research Funding Call 2023 (PI: Raiteri). This work has been partially supported by ICSC – Centro Nazionale di Ricerca in High Performance Computing, Big Data and Quantum Computing, funded by European Union – NextGenerationEU.
\newline
The research has made use of the following \texttt{Python} software packages: \texttt{Matplotlib} (\citealt{Hunter2007}), \texttt{Pandas} (\citealt{McKinney2010}), \texttt{NumPy} (\citealt{vanderwalt2011}), \texttt{SciPy} (\citealt{Virtanen2020}).

\end{acknowledgements}

\bibliography{references.bib}
\bibliographystyle{aa.bst}

\begin{appendix} 
\section{Cross-calibration of light curves}
\label{subsec:agn_lc_calib}
The main challenges when comparing data from different surveys arise from differences in the the detector properties, photometric systems, and the overall data reduction process. The normalised filter transmission curves for SDSS, PS1, and ZTF \textit{gri} bands are reported in Fig. 1 of \cite{Ngeow2019}, whereas Fig. 3 of \cite{Jordi2010} shows the Gaia broad filters. It is clearly evident that while the filters for the ground-based surveys can be adapted via relatively small colour corrections, the Gaia wide field photometry requires more dedicated attention. Indeed, by following \cite{Jordi2010} and the DR3 documentation\footnote{\href{https://gea.esac.esa.int/archive/documentation/GDR3/}{https://gea.esac.esa.int/archive/documentation/GDR3/}}, it is possible to use three wide bands to match the SDSS system with a 4th order polynomial expression:
\begin{equation}
\begin{aligned}
    G - r^{SDSS} &= -0.09837 + 0.08592\ (G_{BP}-G_{RP}) + \\[0.5em] 
    &\quad 0.1907\ (G_{BP}-G_{RP})^2 - 0.1701\ (G_{BP}-G_{RP})^3\\[0.5em] 
    &\quad + 0.02263\ (G_{BP}-G_{RP})^4.
\end{aligned}
\label{eq:Gaia_filt_calib}
\end{equation}
\noindent
A similar equation can be used to go from $G$ to $g^{SDSS}$, but starting from the the same broad band photometry means that it would not add information to the data, as they would be nothing but shifted version of the same magnitude. Hence we can only use the broad filter from Gaia calibrated to either the \textit{g} or \textit{r} SDSS band.

The ZTF Data Products also come with filter-based zero-points ($ZP_f$) and colour-coefficients ($c_f$) for photometric calibration with the PS1 system (\citealt{Masci2019}). The colour correction is defined as polynomial equations:
\begin{equation}
\begin{array}{l}
    g^{PS1}-g^{ZTF} = ZP_g + c_g(g^{PS1}-r^{PS1}), \\[1em] 
    r^{PS1}-r^{ZTF} = ZP_r + c_r(g^{PS1}-r^{PS1}),
\end{array}
\label{eq:ztf_filt_calib}
\end{equation}
\noindent
where the $ZP$ and $c$ coefficients are provided in ZTF DR19, for each epoch. 

\begin{figure}
    \centering
    \includegraphics[width=\linewidth]{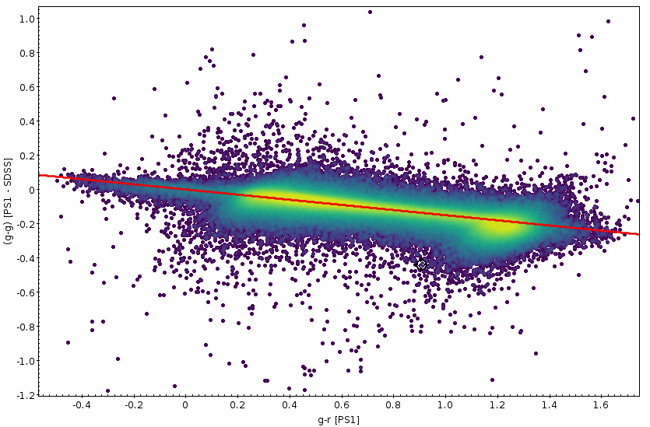}
    \vspace{1mm}
    \includegraphics[width=\linewidth]{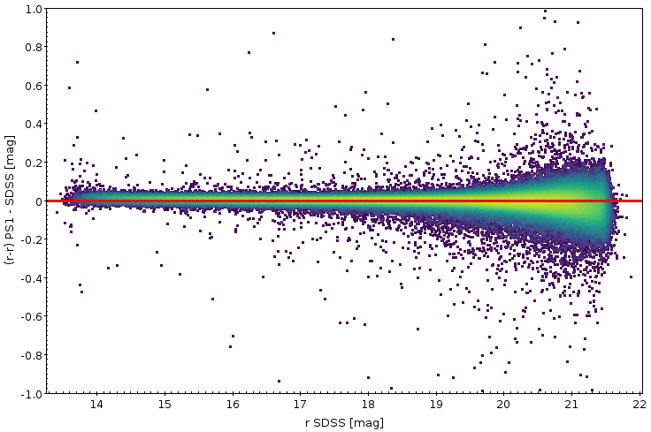}
    \caption{Colour-magnitude diagrams with PS1 and SDSS magnitudes of non-variable stars. The points are colour-coded by density, with yellow indicating regions of higher point density, and darker areas representing lower-density regions. The upper panel shows the linear correction applied to the \textit{g} band, while the lower panel demonstrates that the \textit{r} filters are nearly identical, requiring no correction.}
    \label{fig:color_cal}
\end{figure}

The final step in colour correction is to compare the SDSS and PS1 filters, in order to correct any offsets and standardise all data to a common photometric system. To achieve this, we used a sample of $\sim 5 \times 10^5$ non-variable Stripe-82 stars from \cite{ivezic2007} to create colour-magnitude diagrams with PS1 and SDSS PSF photometry. We found that \textit{r} band photometry is nearly equivalent, at a 1\% level up to 20.5 mag, thus requiring no correction (as also reported in \citealt{Suberlak+21}). The \textit{g} band, instead, presents an offset that we can correct with a linear fit,
\begin{equation}
g^{PS1} - g^{SDSS} = -0.04 - 0.11\ (g^{PS1}-r^{PS1}).
\label{eq:color_corr}
\end{equation}
The colour differences and best-fit results are shown in Fig. \ref{fig:color_cal}. 

As an additional check, we verified the validity of the corrections from ZTF to PS1 system through eqs. \ref{eq:ztf_filt_calib}. Furthermore, we used light curves of non-variable stars to search for any residual biases. 
After a 5-$\sigma$ clipping in magnitude and errors to limit the impact of bad photometry (mostly due to a few erratic PS1 epochs), deviations with respect to the mean value remain of the order of a few sigma. However, it is worth mentioning that the ZTF median error on magnitude is ten times larger than that of SDSS, which is expected because of the lower sensitivity. The same test could not be done for Gaia, due to the lack of light curves for non-variable sources. This is a critical point if our aim is to use the full combined light curve to extract the PSD, but future Gaia data releases could help resolve this issue.

To minimise the uncertainties due to photometric transformations, we limited our analysis to the \textit{r} band for the combined light curves from the four surveys (following \citealt{Suberlak+21}). This way, SDSS and PS1 require no modification, while ZTF and Gaia are corrected as explained above. While the photometric uncertainties on the single light curve might prevent an accurate determination of variability parameters, our binning procedure over discrete time intervals and ensemble PSD analysis should mitigate the impact of small fluctuations, as shown in P24. 

An example of cross-calibrated r-band light curve is shown in Fig. \ref{fig:agn_lc_comb}. Magnitudes are converted into fluxes in units of nJy, following standard procedure for error propagation and using the zero points associated with each survey.

\section{PSD of simulated light curves}
\label{sec:app_b}

\begin{figure*}
    \centering
    \includegraphics[width=0.33\linewidth]{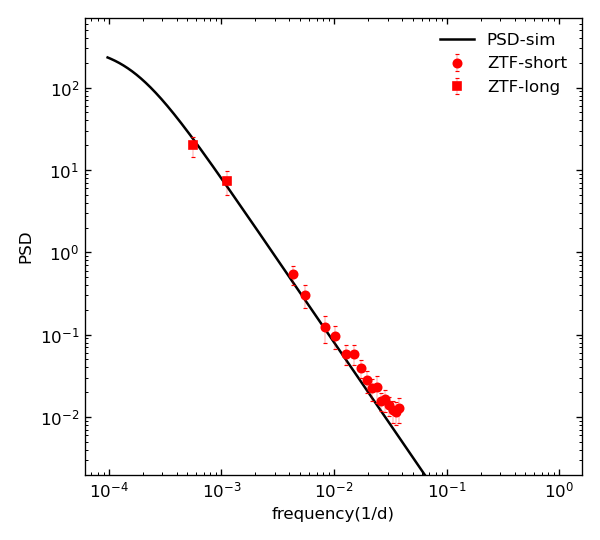}
    \includegraphics[width=0.33\linewidth]{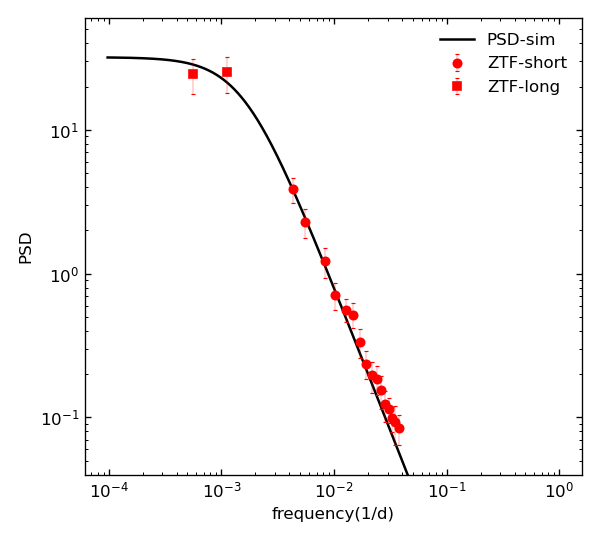}
    \includegraphics[width=0.33\linewidth]{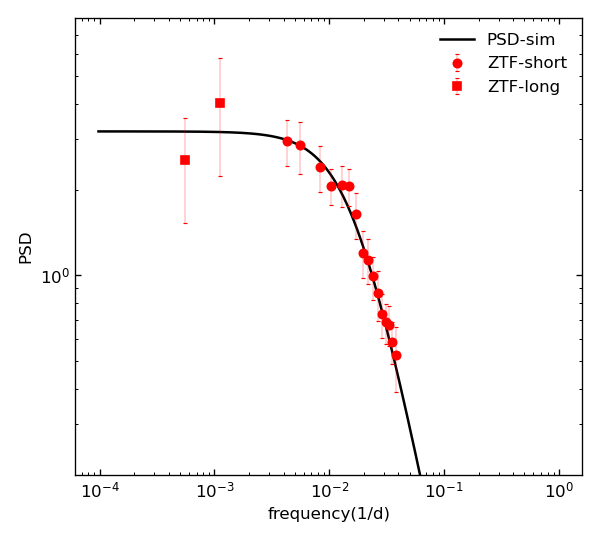}
    \caption{Ensemble power spectra of simulated DRW light curves with the same observing pattern as the ZTF data. Red circles and squares show the measured PSDs at high and low frequencies, respectively. Solid black line refers to the power spectrum associated to a DRW model with $\sigma = 0.28$, and $\log \tau =3$ (left panel), $\log \tau =3$ (middle panel), and $\log \tau =1$ (right panel).}
    \label{fig:sim_int_ztf}
\end{figure*}

To assess potential biases introduced by the linear interpolation in ZTF light curves for the high-frequency PSD, we tested it on simulated data for which we know the intrinsic PSD shape. We used the same approach as P24 (see their Appendix B) to simulate DRW light curves with any given amplitude, $\sigma$, and damping timescale, $\tau$, following the procedure described in \cite{MacLeod10} and \cite{Kovacevic2021}. For each fixed DRW $\tau$ and $\sigma$, we created 69 simulated light curves and we applied the same sampling pattern as the ZTF data from the master sample. We then process the data to measure the PSD exactly as we do in Sect. \ref{sec:ztf_psd}, that is, we bin each seasonal segment with a 12 days window, apply linear interpolation for eventual gaps, and compute the ensemble PSD. The requirements for the interpolation of missing data are very strict: no more than three consecutive gaps and no more than 20\% made up of interpolated points. We stress again that the only reason behind the need for interpolation is that Fourier analysis requires evenly sampled light curves with no zeros or gaps. Otherwise, these zeros will be interpreted as real numbers, introducing very large amplitude features at high frequencies, which will be artificial and seriously affect the estimation of the true, intrinsic PSD of AGN.

Figure \ref{fig:sim_int_ztf} shows examples of power spectra derived from simulated ZTF-like light curves, together with the intrinsic DRW model. To further test the reliability of our analysis, we show both the high-frequency PSD (subject to linear interpolation) and the two low-frequency points estimated as in Sect. \ref{sec:combined1}. For any fixed DRW variability amplitude, we explored different damping timescales to test the accuracy of the reconstructed PSD against different break positions. The figure clearly shows that the recovered PSD is in agreement with the input in any configuration. The slight mismatch at very high-frequencies is probably due to aliasing (i.e. higher non-sampled frequencies folded back to lower frequencies). This is a well know and expected effect, hence, we believe this small offset is not due to interpolation effects. In any case, in the real light curves, the frequency range where it appears is already dominated by the Poisson noise PSD component, which has a much larger amplitude than this mismatch, and it is not affect it by it. Therefore, we do not believe that the linear interpolation scheme we have adopted in this work is affecting the reliability of our PSD analysis.


\section{RBF interpolation}
\label{sec:rbf_int}

In Sect. \ref{sec:comb_int}, we describe our test of a radial basis function (RBF) interpolation method, using the Python SciPy class \texttt{RBFInterpolator}\footnote{\url{https://docs.scipy.org/doc/scipy/reference/generated/scipy.interpolate.RBFInterpolator.html}}. RBFs are scalar functions whose value at any point $x$ can be expressed in terms of the distance from the centre, $r=||x-c||$, where $c$ is the centre of the RBF. For applications with time series, points are the individual observations, while spatial distances are indeed temporal. RBF interpolation consists of expressing a target function $f(x)$ as a linear combination of RBFs, each centred at one of the data points, and polynomials (\citealp{Wahba1990}; \citealp{Fasshauer2007}). A proper combination of such functions can allow us to gather information about missing epochs, by following the overall trend of the light curve. One of the advantages of RBF interpolation over more advanced methods, like Gaussian processes (GPs), is its fast and straightforward implementation, with minimal assumptions on the intrinsic variability model. 

More into detail, for a vector of fluxes \(\mathbf{f}\) measured at times \(\mathbf{t}\), the RBF interpolant function at the new times \(\tilde{\mathbf{t}}\) can be written as\begin{equation}
    \mathbf{f}(\tilde{\mathbf{t}}) = K(\tilde{\mathbf{t}}, \mathbf{t}) \mathbf{a} + P(\tilde{\mathbf{t}}) \mathbf{b},
\label{eq:rbf1}
\end{equation}
\noindent where \(K(\tilde{\mathbf{t}}, \mathbf{t})\) is a matrix of RBFs with centres at \(\mathbf{t}\) evaluated at the times \(\tilde{\mathbf{t}}\), and \(P(\tilde{\mathbf{t}})\) is a matrix of polynomials of any degree. The use of polynomials is optional, and in some cases they may help to preserve global trends or satisfy boundary conditions. The coefficients \(\mathbf{a}\) and \(\mathbf{b}\) are determined by solving the following system of linear equations:
\begin{equation}
    \left( K(\mathbf{t}, \mathbf{t}) + \lambda I \right) \mathbf{a} + P(\mathbf{t}) \mathbf{b} = \mathbf{f},
\label{eq:rbf2}
\end{equation}
\noindent and
\begin{equation}
    P(\mathbf{t})^T \mathbf{a} = 0,
\label{eq:rbf3}
\end{equation}
\noindent where \(\lambda\) is a non-negative smoothing parameter that controls how closely the interpolant fits the data (exact fit for $\lambda=0$), and $I$ is the identity matrix. In the system, Eq. \ref{eq:rbf3} in the system ensures that the RBF terms are independent of the polynomial terms. In other words, it separates the local interpolation from any global trends the polynomial might capture, such as constant shifts or other complex behaviour in the data, while the RBFs handle local variations.

To measure the PSD on long timescales, we tested the RBF interpolation on the combined light curves with different choices of parameters. The best results are found with a multiquadric RBF kernel,
\begin{equation}
  \phi(r)
  \;=\;
  \sqrt{\,1 + \bigl(\varepsilon\,r\bigr)^{2}\,},
\label{eq:rbf4}
\end{equation}
\noindent where $r$ is the distance and $\epsilon$ is the shape parameter controlling the width of the kernel. We used the default shape given by the median pairwise distance between the input times. As a smoothing factor we use $\lambda=0.8$, which balances the fit to the data while avoiding over-fitting the noise. Adding high-order polynomials does not improve the PSD, so we use the default implementation with degree = 0. 

Once the interpolant is built, we can evaluate it on a uniform grid,
\begin{equation}
\tilde{\mathbf{t}} = \bigl[t_{\min},\,t_{\min}+\Delta t,\,\dots,\,t_{\max}\bigr],
\end{equation}
\noindent with a user-defined cadence. We tested values of 36, 180, or 360 days, with shorter cadences resulting in better reconstructed light curves. However, we estimate the final PSD always with a 360 days window, and frequencies shorter than $10^{-3}\ day^{-1}$ are not affected by the RBF interpolation cadence. For each interpolated point, flux errors are calculated using a linear interpolation of the original errors. Additionally, we use a running sigma-clipping over a 36-day window on the light curve to reduce biases from bad photometry, and inject Gaussian noise to the interpolated points to account for measurement uncertainties. 

\begin{figure}
    \includegraphics[width=1\columnwidth]{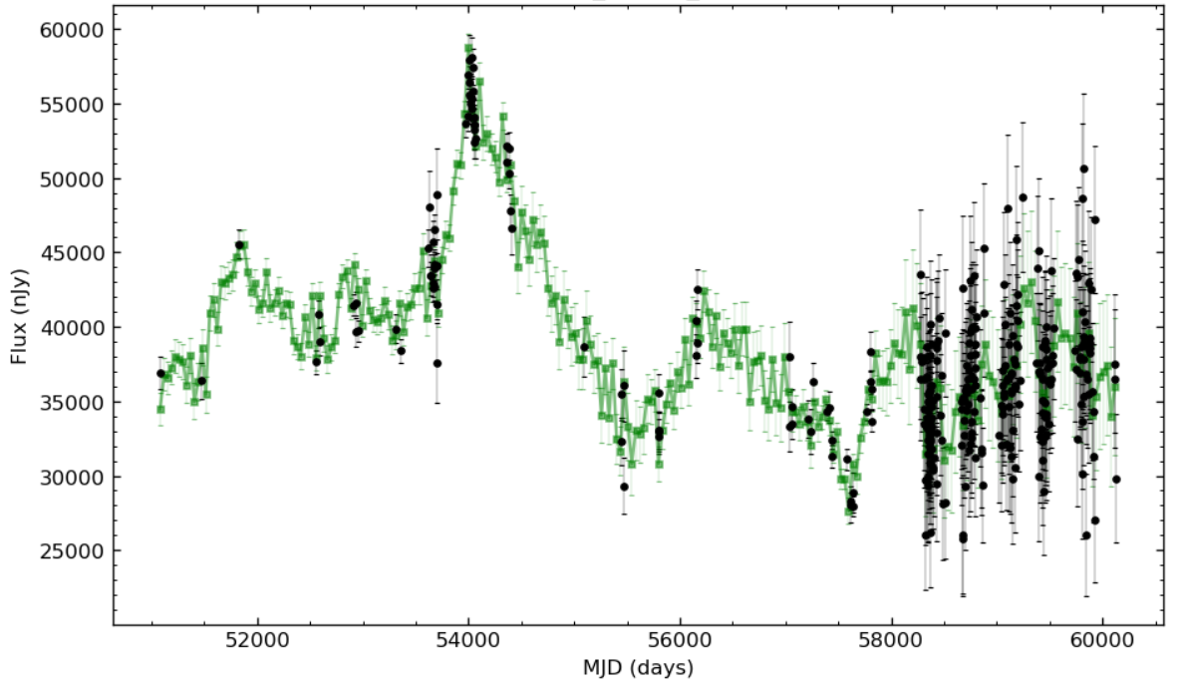}
    \caption{Example of quasar \textit{r}-band light curve combining all the surveys (black points), with the RBF interpolated version over a 36-days window (green points).}
    \label{fig:agn_lc_rbf}
\end{figure}

An example of combined light curve, and the corresponding RBF interpolation with a 36-day cadence is shown in Fig. \ref{fig:agn_lc_rbf}. While the interpolation may be not reliable to reconstruct the short-terms variability (strongly depending on the RBF parameters and the distribution of points), it might be used to fill sporadic gaps on timescales longer than the typical length of an observing season. We used interpolated light curves as described here to measure the PSD at short frequencies shown in Fig. \ref{fig:psd_combined_int}. 

\begin{figure*}
    \centering
    \includegraphics[width=0.33\linewidth]{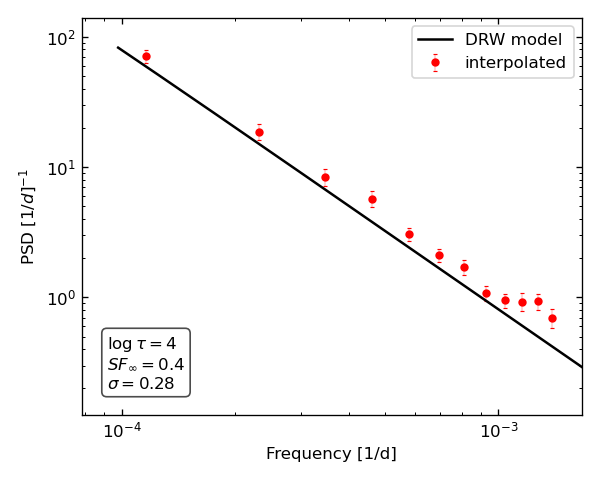}
    \includegraphics[width=0.33\linewidth]{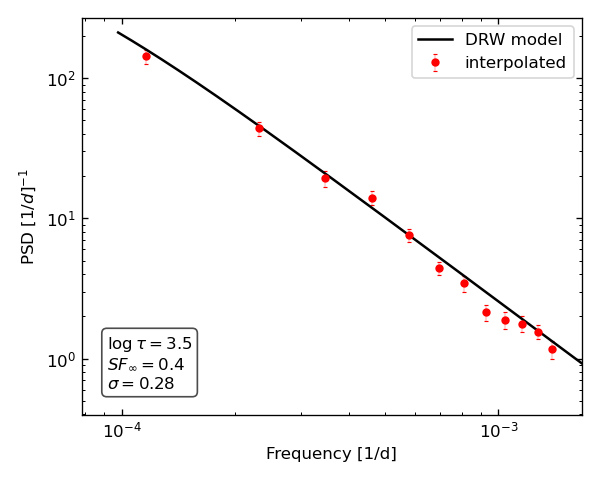}
    \includegraphics[width=0.33\linewidth]{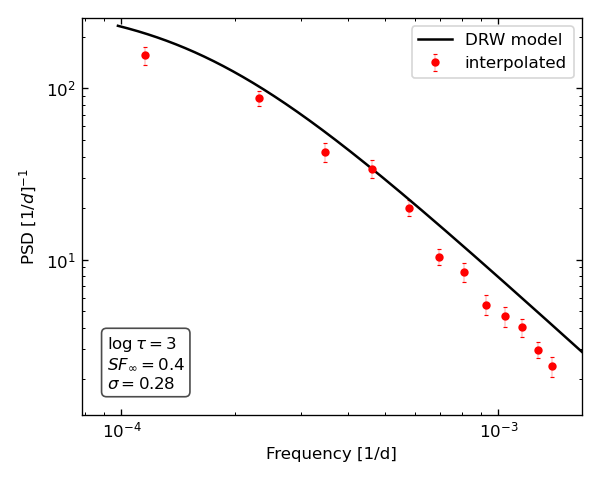}
    \includegraphics[width=0.33\linewidth]{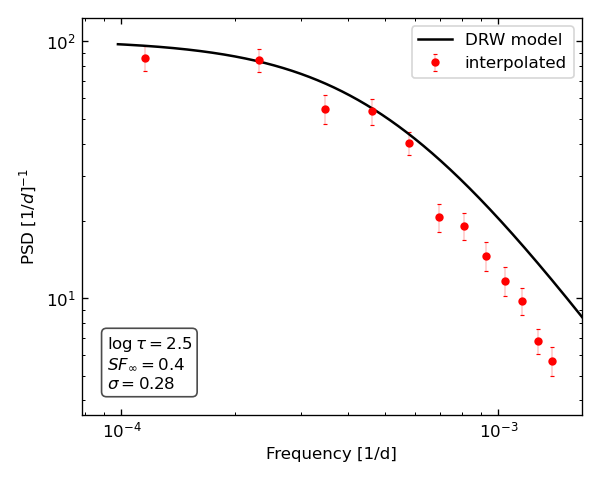}
    \includegraphics[width=0.33\linewidth]{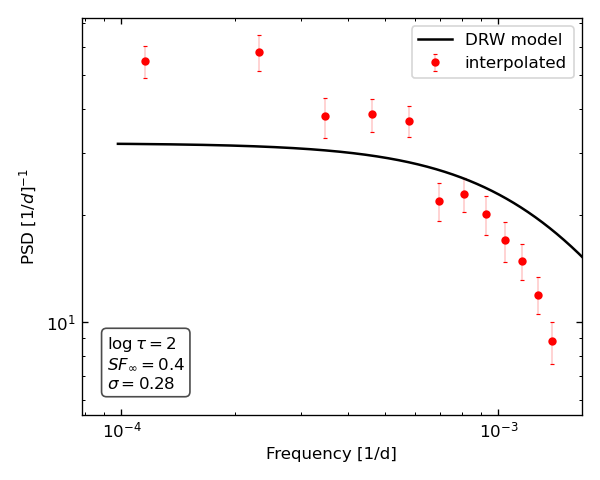}
    \includegraphics[width=0.33\linewidth]{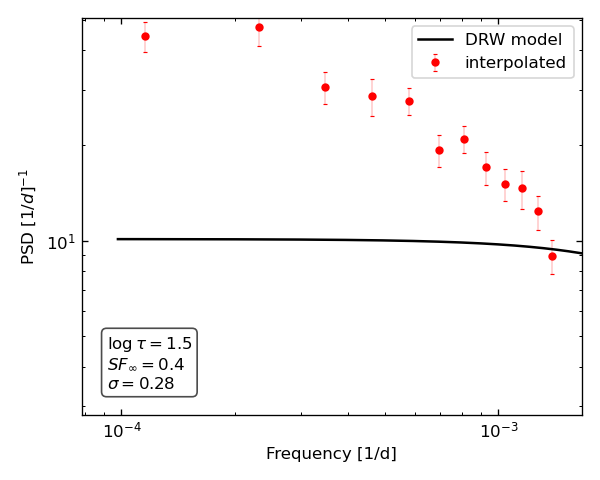}
    \caption{Ensemble power spectra of simulated DRW light curves with the same observing pattern as combined SDSS+PS1+Gaia+ZTF data, interpolated via RBFs. Red points show the measured power spectral densities, while the black solid line refers to the power spectrum associated to the DRW model. Each panel has DRW light curves with different damping timescales, as listed on the figures, probe regions where the interpolation falls before, within or after the break.}
    \label{fig:sim_int}
\end{figure*}

To verify the reliability of the method in reconstructing the power spectrum, we tested it on simulated light curves as described in Appendix \ref{sec:app_b}. For each fixed DRW $\tau$ and $\sigma$, we created 69 simulated light curves and we applied the same sampling pattern as that of the combined SDSS+PS1+Gaia+ZTF data from the master sample. We then interpolated the data with the RBF method and computed the power spectrum as we do for the real data. Figure \ref{fig:sim_int} shows examples of power spectra derived from interpolated light curves, together with the intrinsic DRW model. For a fixed variability amplitude, we explored damping timescales in the range $[10^1-10^4]$ days, to probe regions where the interpolated points are before, within or after the break. The low-frequency part of the reconstructed PSD is generally compatible with the model in the power-law and the early-break regimes (first four panels). However, in the last two cases where the intrinsic PSD flattens, the interpolated light curves show residual power in disagreement with the model. Moreover, also in the cases where they seem to be in agreement, spurious breaks appear in the interpolated power spectra. This clearly shows that our RBF-interpolation cannot be used as a universal interpolation or the reconstruct the unknown PSD shape of a stochastic process, as the outcome depends on the underlying process. For this reasons, we do not include the interpolated points in the analysis, albeit we show here our preliminary attempts and the drawbacks to the interested reader. More extended simulations with other RBF parameters and additional interpolation methods are required, to fully exploit the potential of combined light curves.

\end{appendix}

\end{document}